\documentclass[10pt,superscriptaddress]{revtex4-2}
\usepackage{amssymb}
\usepackage{amssymb}
\usepackage{latexsym}
\usepackage{epsfig}
\usepackage{float}
\usepackage{graphicx,epsfig,color}
\usepackage{graphicx}
\usepackage{subfigure}
\usepackage{amsmath}
\usepackage{hyperref}
\usepackage[sort&compress]{natbib}
\begin{document}
\title{Nonlinear Yang-Mills black holes}
\author{Fatemeh Masoumi Jahromi}
\email{fatemehmasoumi@ph.iut.ac.ir}
\affiliation{Department of Physics, Isfahan University of Technology, Isfahan, 84156-83111, Iran}

\author{Behrouz Mirza}
\email{b.mirza@iut.ac.ir}
\affiliation{Department of Physics, Isfahan University of Technology, Isfahan, 84156-83111, Iran}
 
\author{Fatemeh Naeimipour}
\email{sara.naeimipour1367@gmail.com} 
\affiliation{Department of Physics, Isfahan University of Technology, Isfahan, 84156-83111, Iran}

\author{Soudabe Nasirimoghadam}
\email{snasirimoghadam@sirjantech.ac.ir}
\affiliation{Department of Physics, Sirjan University of Technology, Sirjan, 78137, Iran}

\begin{abstract}
This paper is devoted to investigating the nonlinear non-abelian Yang-Mills black holes. We consider three Born-Infeld, exponential, and logarithmic nonlinear Yang-Mills theories with $SO(n-1)$ and $SO(n-2,1)$ semi-simple groups, which n is the dimension of spacetime, and obtain a new class of nonlinear Yang-Mills (NYM) black hole solutions. Depending on the values of dimension $n$, Yang-Mills charge $e$ and the mass $m$ and nonlinear parameters $\beta$, our solutions can lead to a naked singularity, a black hole with two horizons, an extreme or a Schwarzschild-type black hole. We also investigate the thermodynamic behaviors of the NYM black holes. For small charge values, the NYM solutions may be thermally stable in the canonical ensemble, if we consider an AdS spacetime with spherical $k=+1$ and hyperbolic $k=-1$ coordinates or a flat one with $k=+1$. However, there are no stable regions 
in the grand canonical ensemble in higher dimensions. For the NYM black hole, we observe a reentrant phase transition between large and small black holes in the BI-branch with small $\beta$, which cannot be visible for the nonlinear Reissner-Nordstrom AdS black hole in the higher dimension. For the limit $\beta\rightarrow\infty$, the critical ratio $\frac{P_{c} v_{c}}{T_{c}}$ tends to the constant value $3/8$ for each dimension $n$, while it depends on the dimension for the case of nonlinear electrodynamics black holes.

\end{abstract}

\pacs{04.70.-s, 04.30.-w, 04.50.-h, 04.20.Jb, 04.70.Bw, 04.70.Dy}

\maketitle

\section{Introduction}
The idea of nonlinear electrodynamics has been known as a powerful tool to modify the classical Maxwell theory. The first nonlinear model was introduced by Born and Infeld to remove the self-energy singularity of the point-like particles \cite{Born} by an upper bound imposing on the electric field. They proposed a Lagrangian with a nonlinear parameter $\beta$ that has a dimension of mass and measures the strength of the nonlinearity. As the Born-Infeld (BI) action describes the low energy open superstrings and D-branes dynamics \cite{Fradkin, Tseytlin, Abouelsaood, Bergshoeff, Leigh}, nonlinear electrodynamics can be significant in the framework of string theory. Heisenberg and Euler concluded that vacuum polarization effects are obtained only when the Maxwell equations are substituted by more fundamental $U(1)$ gauge theories of nonlinear electrodynamics \cite{Heisenberg}. Nonlinear electrodynamics reduces to the linear Maxwell theory in the weak field limit. For $\beta=0$, the theory is in the most strongest nonlinear regime, while for large $\beta$, it reduces to the linear-Maxwell theory plus some correction terms proportional to $1/\beta^2$. Born-Infeld Lagrangian could also solve the problems caused by infinite self-energy in the formulation of a quantum theory of electrodynamics \cite{Goenner}. This theory has also corrected shock wave characteristics \cite{Boi} and it cannot predict vacuum birefringence. In recent times, the nonlinear electrodynamics theory has attracted great attention. For instance, the BI action as a specific model naturally appears in D-branes and open superstrings \cite{Leigh}. This theory also plays an effective role to construct a regular black hole solution \cite{Beato} and avoids the singularity problem in the early universe \cite{Bandaos}. It can also influence the critical quantities such as the phase transition point and the gap frequency \cite{Sheykhi0}. In addition to the BI theory, some other types of nonlinear electrodynamics such as the exponential and logarithmic forms \cite{LN, EN} can remove or reduce the singularity of the electric point charge field as well. Within the framework of exponential $U(1)$ gauge theory \cite{soo}, the results indicate a finite value for the total electrostatic field energy, while the electric field at the location of the elementary point charge is not finite. The authors have proved that BI, logarithmic, and exponential $U(1)$ gauge theories can lead to a finite self-energy in arbitrary dimensions \cite{soo}. In Refs. \cite{Alam, Soro}, the general properties of non-linear electrodynamics are reviewed.  \\
\indent The linear abelian Maxwell theory is a subset of a larger class of non-abelian theories such as the Yang-Mills theory. The Maxwell equations can be regarded as the Yang-Mills ones with the gauge group $U(1)$. There are many motivations in order to consider the non-abelian Yang-Mills theory. Spin currents of the ferromagnets correspond to the SU(2) gauge fields in the dual gravitational theory, which open new perspectives for condensed matter systems dual to non-abelian Yang-Mills gauge theories \cite{Yoko}. The Yang-Mills equations may also appear in the low energy limit of some string models. Furthermore, the Yang-Mills theory may also be an important issue in order to characterize the quark confinement by magnetic monopoles and their condensation with a dual superconductor picture. Due to the partial gauge fixing, the abelian projection method explicitly breaks both the local and global gauge symmetry \cite{Hoo,Ezawa}. However, using a new gauge invariant procedure in SU(N) Yang-Mills theory, the gauge-invariant non-abelian magnetic monopole has been successfully introduced \cite{Kondo}. It has also been shown that these non-abelian magnetic monopoles contribute significantly to the confinement of the fundamental quarks in the SU(3) Yang-Mills theory. Moreover, non-abelian excitations like Majorana fermions, can also be used in topologically protected quantum computations \cite{Nayak,Ivanov,Tewari}. Therefore, in order to obtain a broader class of black hole solutions and study the effects of the non-abelian gauge fields on gravity, it is worthwhile to consider the Einstein-Yang-Mills (EYM) theory. Bartnik and McKinnon were the first who found a static and spherically symmetric solitonic solution for the $SU(2)$ gauge group in EYM theory \cite{Bartnik}. Some black hole solutions have also been found in this theory. The black hole solution in EYM theory with SO(3) gauge group has been studied in \cite{Yasskin}, and the colored black holes with SU(2) gauge group have been studied in \cite{Volkov, bizonp}.\\
\indent Naturally, considering a non-abelian theory as a matter field may lead to a set of complicated field equations. In this regard, most of the black hole solutions in EYM theory are numerical \cite{Deveci, Van, Okuyama0}. However, some authors have considered some particular ansatzes in order to find analytical solutions \cite{Okuyama, Radu, Brihaye}. For the first time, Yasskin could achieve the black hole solution for the EYM equations using the Wu-Yang ansatz \cite{Yasskin}.
 Wu-Yang \cite{Wu} is one of the ansatzes which has been used in a lot of papers to obtain black hole solutions \cite{Mazharimousavi, Dehghani,Deh00,mazhari1,mazhari2,mazhari3,mazhari4}. Topological black hole in Gauss-Bonnet-Yang-Mills gravity has been studied in Ref. \cite{Bostani}. Thermodynamic behaviors of the EYM black hole in the presence of massive gravity have been also probed \cite{Nemati}. In Ref. \cite{Stetsko00}, static spherically symmetric Einstein-Maxwell-Yang-Mills-dilaton black hole and the related thermodynamics have been investigated. Recently, we have reached the Yang-Mills black hole solutions in the modified quasitopological gravity \cite{Mir}. A set of numeric non-abelian Einstein-Born-Infeld solutions has been also studied in Ref. \cite{Dyadichev}. Some have already obtained the analytic solutions for the non-abelian BI theory in the presence of the Einstein and Gauss-Bonnet gravities. \cite{Mazharimousavii}. Some studies of different non-abelian Yang-Mills black holes were done in Refs. \cite{Maz,Zhang1,Ste1,Ste2,Ali} as well.
Now, we aim to use the Wu-Yang ansatz and access a vast $n$-dimensional Yang-Mills black hole solutions in the presence of three Born-Infeld, exponential, and logarithmic nonlinear forms. We also investigate the physical and thermodynamic properties such as thermal stability and critical behavior, and also the dynamical stability of the obtained solutions. \\
\indent This paper is organized as follows: In Sec. \ref{Field}, we define the main structure of the Einstein gravity coupled to the nonlinear non-abelian Yang-Mills gauge fields and then obtain the related black hole solutions. We also investigate the physical structures of the solutions in Sec. \ref{Phys}. Thermodynamic properties and thermal stability of the obtained black hole solutions are probed in Secs. \ref{thermo} and \ref{stability}, respectively. In Sec. \ref{P-V}, we study the critical behavior of the related black holes in the extended phase space. We also study the dynamical stability of the NYM black holes in Sec. VI E. Finally, we have a conclusion of the whole paper in Sec. \ref{result}.  
\section{Main structure of the NYM Black hole solutions}\label{Field}
The theory of the $n$-dimensional ($n\geq4$) nonlinear-Yang-Mills (NYM) black hole solutions originate from the action  
\begin{eqnarray}\label{action}
I=\frac{1}{16\pi}\int_{\mathcal{M}}{d^{n}x \sqrt{-g}(R-2\Lambda+L(F))},
\end{eqnarray}
where $R$ and $\Lambda$ are respectively the Ricci scalar and the cosmological constant. 
We classify the nonlinear Lagrangian $L(F)$ for three Born-Infeld, exponential and logarithmic cases \cite{Shey3}
\begin{eqnarray}\label{22}
L(F)=\left\{
\begin{array}{ll}
$$4\beta^2\bigg[1-\sqrt{1+\frac{F^2}{2\beta^2}}\bigg]$$,\quad\quad\quad \quad\quad \,\, \ {BI}\quad &  \\ \\
$$4\beta^2\bigg[e^{-\frac{F^2}{4\beta^2}}-1\bigg]$$,\quad\quad\quad \quad\quad\,\,\,\,\,\,\,\,\,\,\,  \ {EN}\quad &  \\ \\
$$-8\beta^2 \mathrm{ln}\bigg[1+\frac{F^2}{8\beta^2}\bigg]$$, \quad\quad\quad\quad\quad\quad\,  \ {LN}\quad &
\end{array}
\right.
\end{eqnarray}
where we have abbreviated the exponential and logarithmic nonlinear cases to EN and LN, respectively.
If we consider a gauge group with N-parameters, so
\begin{eqnarray}\label{gam}
F^2=\gamma_{ab}F_{\mu\nu}^{(a)}F^{(b)\mu\nu},\,\,\,\,\gamma_{ab}\equiv-\frac{\Gamma_{ab}}{|\mathrm {det} \Gamma_{ab}|^{1/N}},
\end{eqnarray}
where $\Gamma_{ab}=C_{ad}^{c}C_{bc}^{d}$ is the metric tensor of the gauge group and $\mathrm {det} \Gamma_{ab}$ is the related determinant. The indices $a,b,c$ take values from 1 to $N$ and $C^{a}_{bc}$'s are the structure constants of the gauge group theory. The gauge field tensor is defined as 
\begin{eqnarray}
F_{\mu\nu}^{(a)}=\partial_{\mu}A_{\nu}^{(a)}-\partial_{\nu}A_{\mu}^{(a)}+\frac{1}{e}C^{a}_{bc}A_{\mu}^{(b)}A_{\nu}^{(c)}, 
\end{eqnarray}
where $e$ is the coupling constant of the non-abelian theory and $A_{\mu}^{(a)}$'s represent the gauge potentials.
For simplicity, we use the redefinition $L(F)= \zeta_{1} \beta^2\mathcal{L}(Y)$, where
\begin{eqnarray}
\mathcal{L}(Y)=\left\{
\begin{array}{ll}
$$1-\sqrt{1+Y}$$,\quad\quad\quad  \ {BI}\quad &  \\ \\
$$e^{-Y}-1$$,\quad\quad\quad \,\,\,\,\,\,\,\,\, \ {EN}\quad &  \\ \\
$$\mathrm{ln}(1+Y)$$,\quad\quad\quad\quad  \ {LN}\quad &
\end{array}
\right.
\end{eqnarray}
and $\zeta_{1}=+4,+4,-8$ and $Y=\frac{F^2}{2\beta^2},\frac{F^2}{4\beta^2},\frac{F^2}{8\beta^2}$ are described for BI, EN and LN Yang-Mills theories, respectively. If we vary the action \eqref{action} with respect to the metric $g_{\mu\nu}$ and the gauge potential $A_{\mu}^{(a)}$, then the gravitational and gauge field equations are obtained as follows
\begin{eqnarray}\label{Equ}
R_{\mu\nu}=\frac{2\Lambda}{n-2} g_{\mu\nu}+ \zeta_{2}\gamma_{ab}\partial_{Y}\mathcal{L}(Y)F_{\mu}^{(a)\lambda}F_{\nu\lambda}^{(b)}+\frac{ \zeta_{3}\beta^2}{n-2} [2Y\partial_{Y}\mathcal{L}(Y)-\mathcal{L}(Y)]g_{\mu\nu},
\end{eqnarray}
\begin{eqnarray}\label{elect}
\nabla_{\nu}(\partial_{Y}\mathcal{L}(Y)F^{(a)\mu\nu})=\frac{1}{e}\partial_{Y}\mathcal{L}(Y) C_{bc}^{a}A_{\nu}^{(b)}F^{(c)\nu\mu},
\end{eqnarray}
where $\zeta_{2}=-4,-2,+2$ and $\zeta_{3}=+4,+4,-8$ have been specified for $BI$, $EN$ and $LN$ theories, respectively. To obtain a set of static and
spherically symmetric non-abelian black hole solutions with spherical and hyperbolic horizons, we consider the following metric 
\begin{eqnarray}\label{metric}
ds^2=-f(r)dt^2+\frac{dr^2}{f(r)}+r^2 d\Omega_{k}^{2},
\end{eqnarray}
where $ d\Omega_{k}^{2}$ denotes the line element of an $(n-2)$-dimensional hypersurface $\Sigma$ with the constant curvature $(n-2)(n-3)k$. It is determined as 
\begin{eqnarray}
d\Omega_{k}^{2}=d\theta ^2+k^{-1}\mathrm{sin}^{2}(\sqrt{k}\theta)\bigg(d\phi_{1} ^2+\sum_{i=2}^{n-3}\Pi_{j=1}^{i-1}\mathrm{sin}^2 \phi_{j} d\phi_{i}^2\bigg),
\end{eqnarray}
where $k=1,-1$ are devoted to the spherical and hyperbolic geometries, respectively, and $\theta\in[0,\frac{\pi}{2}]$. If we introduce the coordinates 
$x_{i}$'s 
\begin{eqnarray}
&&x_{1}=\frac{r}{\sqrt{k}}\, \mathrm{sin}(\sqrt{k}\,\theta)\,\Pi_{j=1}^{n-3}\,\mathrm{sin}\,\phi_{j},\nonumber\\
&&x_{i}=\frac{r}{\sqrt{k}}\, \mathrm{sin}(\sqrt{k}\,\theta)\,\mathrm{cos}\,\phi_{n-i-1}\,\Pi_{j=1}^{n-i-2}\,\mathrm{sin}(\phi_{j})\,\,\,\,,\,\,\,i=2, ..., n-2\nonumber\\
&&x_{n-1}= r\, \mathrm{cos}\,(\sqrt{k}\,\theta),
\end{eqnarray}
and employ the Wu-Yang ansatz, then the gauge potentials are obtained from
\begin{eqnarray}\label{gauge}
A^{(a)}&=&\frac{e}{r^2}(x_{a}dx_{n-1}-x_{n-1}dx_{a})\,\,\,\mathrm{for}\,\,\, a=1,...,n-2\nonumber\\
A^{(b)}&=&\frac{e}{r^2}(x_{i}dx_{j}-x_{j}dx_{i})\,\,\, \mathrm{for}\,\,\,i=1,...,n-3\,,\, j=2,...,n-2, \,\mathrm{and}\,i<j  
\end{eqnarray} 
where $b$ goes from $(n-1)$ to $(n-1)(n-2)/2$. The Lie algebra of the gauge potentials with $k=+1$ and $-1$ in Eq. \eqref{gauge} is isomorphic to $SO(n-1)$ and $SO(n-2,1)$ gauge groups, respectively. It should be noted that $n$ is equal to the spacetime dimension. We redefine $\gamma_{ab}$ in Eq. \eqref{gam} as 
\begin{eqnarray}
\gamma_{ab}=\epsilon_{a}\delta_{ab}, \,\,\,\,\,\,\mathrm{no\,\, sum\,\, on\,\, a},
\end{eqnarray}
where for $SO(n-1)$ gauge group, 
\begin{eqnarray}
\epsilon_{a}=1\,\,\,\,, \,\,\,\mathrm{for}\, a=1\,\,\,,...,\frac{(n-1)(n-2)}{2}, 
\end{eqnarray}
and for $SO(n-2,1)$ gauge group
\begin{eqnarray}
\epsilon_{a}=\left\{
\begin{array}{ll}
$$-1$$\quad\quad\quad  \ {1\leq a \leq n-2}\quad &  \\ \\
$$1$$\quad\quad\quad \,\,\,\,\,\,  \ {n-1\leq a \leq \frac{(n-1)(n-2)}{2}}.\quad & 
\end{array}
\right.
\end{eqnarray}
\indent For a better understanding, we have written the gauge potentials of the groups $SO(3)$, $SO(2,1)$, $SO(4)$ and $SO(3,1)$ in appendix \eqref{app}.
The gauge potentials \eqref{gauge} can satisfy the gauge field equation \eqref{elect}. So, if we substitute these gauge potentials \eqref{gauge} and the metric \eqref{metric} in Eq. \eqref{Equ}, we can obtain the NYM black hole solutions
\begin{eqnarray}\label{fff}
f(r)&=& k-\frac{m}{r^{n-3}}-\frac{2\Lambda r^2}{(n-1)(n-2)}\nonumber\\
+&&\left\{
\begin{array}{ll}
$$\frac{4\beta^2 r^2}{(n-1)(n-2)}\bigg[1-\frac{n-1}{r^{n-1}}\int r^{n-2}\sqrt{1+\frac{\eta}{2}}dr\bigg]$$,\quad\quad\quad\quad\quad\quad\quad  \ {BI}\quad &  \\ \\
$$-\frac{4\beta^2 r^2}{(n-1)(n-2)}\bigg[1-\frac{n-1}{r^{n-1}}\int r^{n-2}\mathrm{exp}\big({-\frac{\eta}{4}}\big)dr\bigg]$$,\quad\quad\,\,\,\,\quad\quad \quad  \ {EN}\quad &  \\ \\
$$-\frac{8\beta^2}{(n-2)r^{n-3}}\int r^{n-2}\mathrm{ln}[1+\frac{\eta}{8}]dr$$,\quad\quad\quad\quad\quad\quad\quad\quad\quad\quad\quad  \ {LN}\quad &
\end{array}
\right.
\end{eqnarray}
where $\eta=\frac{(n-2)(n-3)e^2}{\beta^2 r^4}$. The parameter $m$ is an integration constant relating to the mass of the NYM black hole. It should be noted that for $n=4z+1$ where $z\in N$, the metric function in Eq. \eqref{fff} can be written in terms of elementary functions. We have shown the function $f(r)$ for $n=5$ and $n=9$ dimensions in appendix \eqref{f1}. For $n\neq4z+1$, $f(r)$ may be written as
\begin{eqnarray}\label{ff}
f(r)&=& k-\frac{m}{r^{n-3}}-\frac{2\Lambda r^2}{(n-1)(n-2)}\nonumber\\
+&&\left\{
\begin{array}{ll}
$$\frac{4\beta^2 r^2}{(n-1)(n-2)}\big[1-{}_2F_{1}\big(\big[\frac{-1}{2},\frac{1-n}{4}\big]\,,\big[\frac{5-n}{4}\big]\,,-\frac{\eta}{2}\big)\big]$$,\quad\quad\quad\quad\quad\quad\quad\quad\quad\,\,\,\,\,  \ {BI}\quad &  \\ \\
$$-\frac{4\beta^2 r^2}{(n-1)(n-2)}\big[1-{}_2F_{1}\big(\big[\frac{1-n}{4}\big]\,,\big[\frac{5-n}{4}\big]\,,-\frac{\eta}{4}\big)\big]$$,\quad\quad\quad\quad \quad\quad\quad\quad\quad\quad \,\,\,\, \ {EN}\quad &  \\ \\
$$-\frac{8\beta^2 r^2}{(n-1)(n-2)}\mathrm{ln}\big[1+\frac{\eta}{8}\big]-\frac{4(n-3)e^2}{(n-1)(n-5)r^2}{}_{2}F_{1}([1,\frac{5-n}{4}]\,,[\frac{9-n}{4}]\,,-\frac{\eta}{8})$$,\quad\quad  \ {LN}\quad &
\end{array}
\right.
\end{eqnarray}
 Eq. \eqref{ff} shows that there is an equivalence between the four-dimensional NYM black hole solutions with $SO(3)$ and $SO(2,1)$ gauge groups and a set of topological black hole solutions with $k=1$ and $k=-1$ in nonlinear electrodynamics theory \cite{Shey3,Cai0}.
Therefore, we can deduce that there is a transformation between the non-abelian gauge
fields and a set of abelian ones in $n=4$ which satisfies the Yasskin theory \cite{Yasskin}. However, for $n>4$, we achieve a new class of solutions for the NYM black hole which is different from the nonlinear electrodynamics one\cite{Shey3, Cai0}. So, the NYM solutions with $n>4$ do not respect the Yasskin theorem. \\
\indent For large $\beta$, we assume that all three types of the metric functions in Eq. \eqref{fff} reduce to the Einstein-Yang-Mills black hole solution as follows
\begin{eqnarray}\label{f}
	f(r)&=& k-\frac{m}{r^{n-3}}-\frac{2\Lambda r^2}{(n-1)(n-2)}
	+\left\{
	\begin{array}{ll}
		$$-\frac{(n-3)e^2}{(n-5)r^{2}}+\mathcal{O}\big(\frac{1}{\beta^2}\big)$$,\quad\quad\quad  \ {n\neq 5}\quad &  \\ \\
		$$-\frac{2e^2\mathrm{ln}(r/r_{0})}{r^{2}}+\mathcal{O}\big(\frac{1}{\beta^2}\big)$$,\quad \quad  \ {n=5}\quad &  
	\end{array}
	\right.
\end{eqnarray}
 We choose $r_{0}=1$ for simplicity. 
\section{Physical behaviors of the NYM black hole solutions}\label{Phys}
\begin{figure}
\centering
\subfigure[\,$ n=5 $ and $ \beta=3 $]{\includegraphics[scale=0.27]{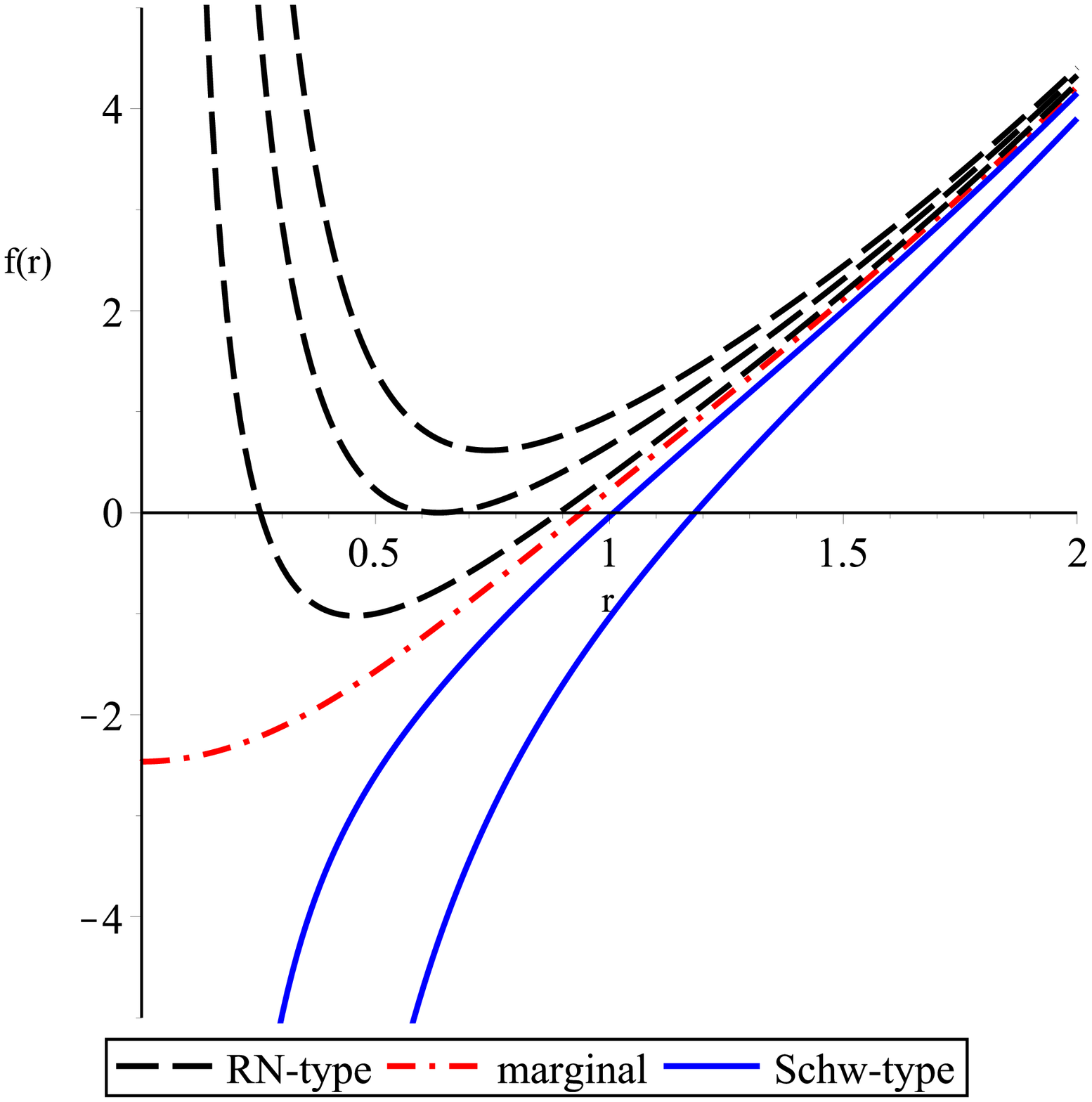}\label{Fig1a}}\hspace*{.2cm}
\subfigure[\,$ n=6 $ and $ \beta=3 $]{\includegraphics[scale=0.27]{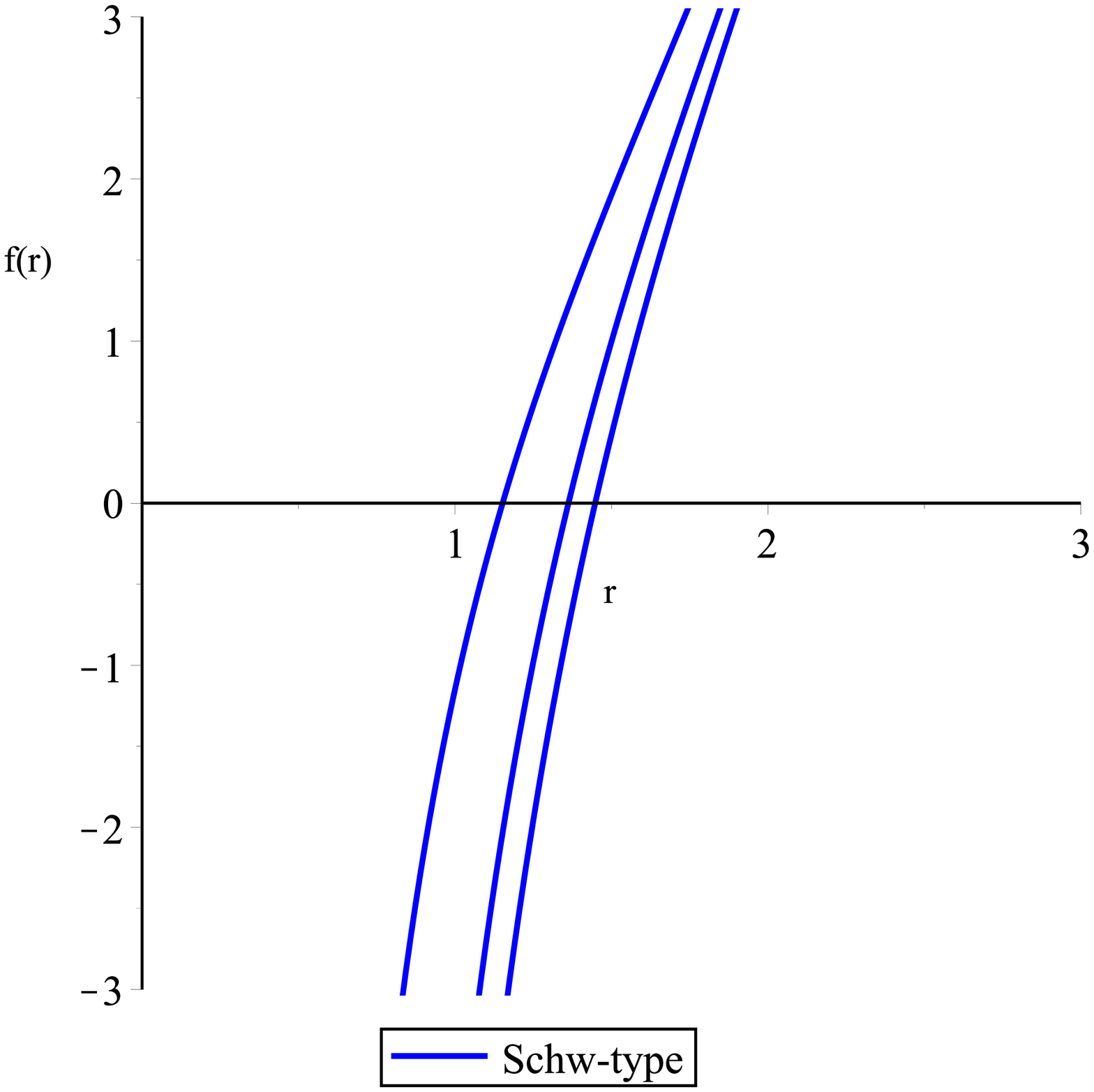}\label{Fig1b}}\caption{Horizon structure for AdS-BIYM black hole solutions in 5 and 6 dimensions. In 5 dimensions, we have the marginal case with red dash-dot line (for $m=A_{5}$), the Schw-type case with blue solid lines (for $m>A_{5}$) and the RN-type case with black dash lines (for $m<A_{5}$). The black dash RN-type lines are defined for the naked singularity, extremal black hole and a black hole with two horizons from top to down . In 6 dimensions, there is only the Schw-type black hole. We have set $ e=1 $, $ k=1 $ and $ l=1 $.}\label{Fig1}
\end{figure}

\begin{figure}
\centering
\subfigure[\,LNYM theory, $ n=5 $ and $ \beta=3 $.]{\includegraphics[scale=0.28]{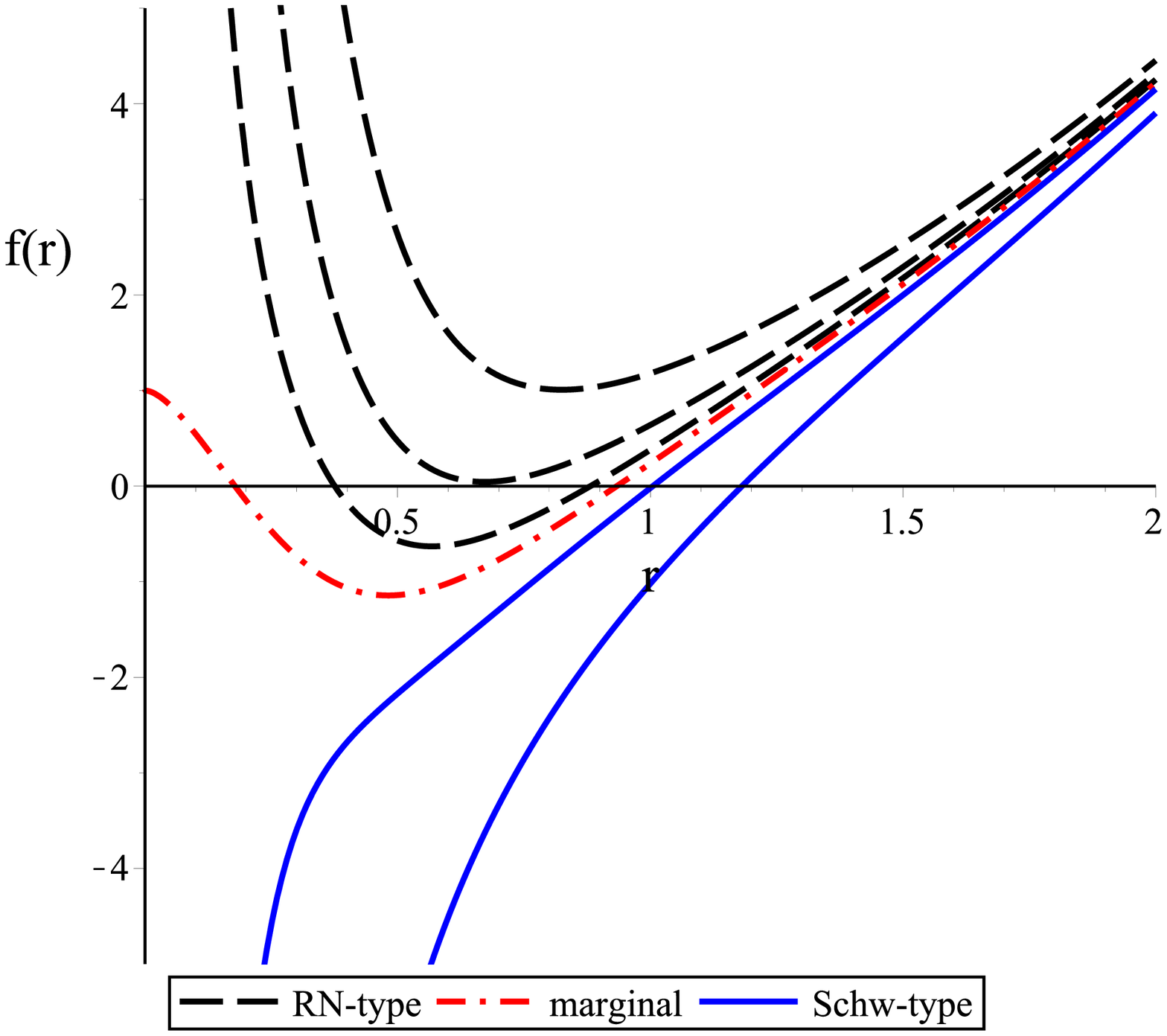}\label{Fig2a}}\hspace*{.2cm}
\subfigure[\,ENYM theory, $ n=5 $ and $\beta$=2.5.]{\includegraphics[scale=0.28]{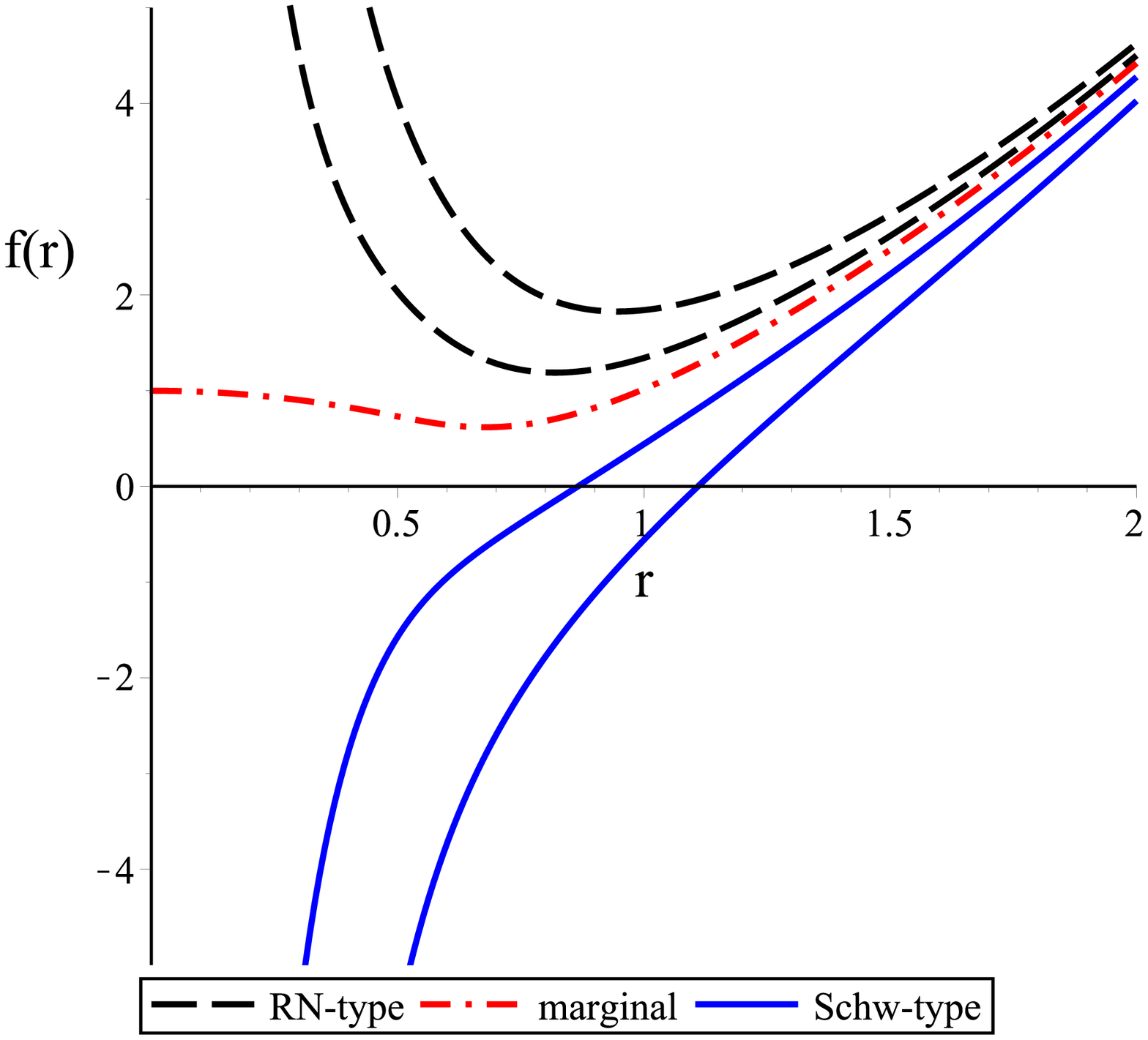}\label{Fig2b}}\caption{ Horizon structure for AdS-LNYM and AdS-ENYM black hole solutions in 5 dimensions, respectively. We have set $ e=1 $, $ k=1 $ and $ l=1 $. }\label{Fig2}
\end{figure}

In this section, we aim to study the physical structures of the NYM black hole solutions. If we calculate the Kretschmann scalar, $R_{abcd}R^{abcd}$, it goes to infinity as $r\rightarrow0$. Therefore, we can deduce an essential singularity located at $r=0$ for the NYM black holes.\\
\indent In the previous section, we concluded that the NYM and nonlinear electrodynamics black hole solutions are the same in $n=4$. In Refs. \cite{Kubiznak, zou, Fernando1, Ams}, the authors have discussed the structure of the nonlinear electrodynamics black holes horizon in $n=4$ and higher dimensions. To find the possible horizons, we investigate the behavior of the metric function $f(r)$ near $r=0$. We have shown the horizon structure for the Born-Infeld Yang-Mills (BIYM) black hole in $n=5,6$ and for $k=1$ in Fig. (\ref{Fig1}). Considering $ k=1 $, and Solving Eq. \eqref{fff} for $ n=5 $ the metric function is
\begin{eqnarray}\label{f5c}
f(r)= 1-\frac{m}{r^2}-\frac{\Lambda r^2}{6}+\frac{\beta^2 r^2}{3}\biggr[1-\sqrt{1+\frac{\eta}{2}}\,\biggr]-\frac{e^2}{r^2}\rm \ln\biggr[r^2(1+\sqrt{1+\frac{\eta}{2}})\biggr]-\frac{4\,\beta^2\,C_{5}}{3\,r^2},
\end{eqnarray}
where $ C_{5} $ is the integration constant for $ n=5 $, which is related to the integral in Eq. \eqref{fff}. The integral in Eq. \eqref{fff} is indefinite, and so we have considered the integration constant. We assume that the expansion of $f(r)$ at large values of $\beta$ $ (\beta\rightarrow \infty) $ reduces to the Yang-Mills (YM) solution in Eq. \eqref{f} and so we find 
\begin{eqnarray}\label{c5}
C_{5}=-\frac{3\,e^2}{8\, \beta^2}\big(1+2\ln(2)\big).	
\end{eqnarray}
Now, the expansion of $ f(r) $ close to $r=0$  in Eq. \eqref{f5c} takes the following form
\begin{eqnarray}\label{bast}
	f(r)&=& 1-\frac{m-A_{5}}{r^{2}}-\frac{2\sqrt{3}}{3}\beta e+\mathcal{O}(r),\,\,\,\,\,A_{5}=\frac{1}{2}e^2\big(1+\mathrm{ln}\big(\frac{4\beta^2}{3e^2}\big)\big)\,\,\,\,\mathrm{for}\,\,\,n=5,
\end{eqnarray}
 where $A_{5}$ is the 'marginal' mass for $n=5$, which depends on the values of parameters $\beta$ and $e$. As we observe in Fig. (\ref{Fig1a}), independent of the parameters, $f(r)$ goes to $\infty$ as $r\rightarrow\infty$. However, for $r\rightarrow 0$, we have the following cases:\\
For the marginal case which is characterized by $m=A_{5}$, the function $f(r)$ in Eq. \eqref{bast} has a finite value at $r=0$ which is
\begin{eqnarray}\label{marg}
f(r)&=& 1-\frac{2\sqrt{3}}{3}\beta e\,\,\,\,\mathrm{for}\,\,\,n=5.
\end{eqnarray}
For $m>A_{5}$, the function $f(r)$ goes to $-\infty$ as $ r\rightarrow0$. Therefore, there is just one horizon and the AdS-BIYM black hole behavior is analogous to the Schwarzschild black hole (we abbreviate it to Schw-type).\\
For $m<A_{5}$, the solution goes to $\infty$ at the limit $r\rightarrow 0$. Thus, the BIYM black hole has a similar behavior like the 'Reissner-Nordstr\"{o}m' black hole (we abbreviate it to RN-type). In this case, the black hole may have zero (naked singularity), one (extremal black hole) or two horizons. Horizons ($r_{+}$) are the roots of the equation $f(r_{+})=0$. If the finite value of the metric function in Eq. \eqref{marg} is positive,(i.e., for $ \beta e<\frac{\sqrt{3}}{2} $), then the solution with $m<A_{5}$ leads to a naked singularity and so the only solution is the Swch-type with $m>A_{5}$. However, when the finite value in Eq. \eqref{marg} is negative ($\beta e>\frac{\sqrt{3}}{2}$), the solution with $m<A_{5}$ can describe a black hole with horizons for $m_{ex}<m<A_{5}$. The parameter $m_{ex}$ is the mass of the extremal black hole, which is determined from the conditions $f(r=r_{ex})=0 $ and $ f^{'}(r=r_{ex})=0 $. For $n=5$, it is given by
\begin{eqnarray}
\left(\frac{4}{l^2}+\frac{4\beta^2}{3}\right)r_{ex}^3+\left(2+\frac{4\beta}{3}\sqrt{\beta^2 r_{ex}^2+3 e^2}\right)r_{ex}=0,\,\,\,\,\,\,\, \mathrm{for}\,\,\,{n=5}
\end{eqnarray}
 We have probed the horizon structure of the BIYM black hole for $n=6$ In Fig. (\ref{Fig1b}). For $ n=6 $, the expansion of the metric function in Eq. \eqref{fff} around $ r=0 $ becomes
\begin{eqnarray}\label{bast2}
f(r)&=& 1-\frac{m-A_{6}}{r^{3}}-\frac{\sqrt{6}}{3}\beta e+\mathcal{O}(r),\,\,\,\,A_{6}=-\frac{12}{5}\sqrt[4]{\frac{6}{\pi^2 \beta^2}}e^{5/2}\,\Gamma\big({\frac{3}{4}}\big)^2,\,\,\, \mathrm{for}\,\,\,{n=6}
\end{eqnarray}
This shows that the marginal case in $n=6$ happens only for $m<0$. Therefore, the six-dimensional BIYM black hole has only one horizon when $m>0$, which is the Schw-type. This behavior is a general feature of the nonlinear Yang-Mills black holes in some higher dimensions that the marginal mass is negative and there can be only one horizon and thus the Schw-type is the only solution.\\
\indent In Figs. (\ref{Fig2a}) and (\ref{Fig2b}), we have investigated the horizon structures of the exponential and logarithmic nonlinear Yang-Mills black holes (we abbreviate them to ENYM and LNYM, respectively) in five dimensions. We observe the same behavior as the BIYM case when $r\rightarrow 0$. Expanding the metric function, one can examine the horizon structure near the origin for ENYM and LNYM cases in the same way as the BIYM case. The expansions of the function $ f(r) $ in Eq. \eqref{fff} around $ r=0 $ for the ENYM and LNYM cases in $ n=5 $ and for $ k=1 $ are given by
\begin{eqnarray}\label{faroundzeroEN}
	f(r)&=& 1-\frac{m-A_{5}}{r^{2}}+\mathcal{O}(r),\,\,\,\,\,A_{5}=\frac{1}{2}e^2\big(1-\gamma+\mathrm{ln}\big(\frac{2\beta^2}{3e^2}\big)\big)\,\,\,\,\mathrm{for}\,\,\,n=5,
\end{eqnarray}
and
\begin{eqnarray}\label{faroundzeroEN}
	f(r)&=& 1-\frac{m-A_{5}}{r^{2}}+\mathcal{O}(r),\,\,\,\,\,A_{5}=\frac{1}{2}e^2\big(1+\mathrm{ln}\big(\frac{4\beta^2}{3e^2}\big)\big)\,\,\,\,\mathrm{for}\,\,\,n=5,
\end{eqnarray}
respectively, where $ \gamma $ is Euler-Mascheroni constant. As we observe in Figs. (\ref{Fig2a}) and (\ref{Fig2b}) when $ m=A_{5} $ we have $ f(r)=1$. For $ m>A_{5} $ the behavior of the metric function is Schw-type. For $ m<A_{5} $ depending on the parameters $\beta$ and $ e $, we may have zero(naked singularity), one(extremal black hole) or two horizons. In fact, the marginal mass, which depends on both $ e $ and $\beta$, is a boundary(or a margin) between two qualitatively different kinds of solutions shown in Fig. (\ref{Fig1a}) and Fig. (\ref{Fig2}). If the constant of integration $ m $ is larger than the marginal mass then the exact solution in Eq. \eqref{fff} has only one horizon that is similar to the Schwarzchild black hole behavior, despite the fact that the black hole is charged. If $ m $ is smaller than the marginal mass then the exact solution \eqref{fff} has two horizons, which is the same as Reissner Nordstr\"{o}m behavior. The marginal mass forms the boundary between these two cases for $ n=5 $ and some higher dimensions. \\
\indent We can also investigate the horizon structure of the solutions in higher dimensions. In general, the expansion of the metric function $ f(r) $ around $ r=0 $  in Eq. \eqref{fff} for the BIYM case is given by 
\begin{eqnarray}\label{faroundzero}
	f(r)&=& k-\frac{m}{r^{n-3}}+\frac{4\beta^2\,C_{n}}{(n-2)\,r^{n-3}}-\frac{2\sqrt{2(n-2)(n-3)}}{(n-2)(n-3)}\beta e+\mathcal{O}(r),
\end{eqnarray}
and for the ENYM and LNYM cases are
\begin{eqnarray}\label{faroundzeroEN1}
	f(r)&=& k-\frac{m}{r^{n-3}}+\frac{4\beta^2\,C_{n}}{(n-2)\,r^{n-3}}+\mathcal{O}(r),
\end{eqnarray}
and 
\begin{eqnarray}\label{faroundzeroLN1}
	f(r)&=& k-\frac{m}{r^{n-3}}-\frac{8\beta^2\,C_{n}}{(n-2)\,r^{n-3}}+\mathcal{O}(r),
\end{eqnarray}
respectively, where $ C_{n} $ is the integration constant for dimension n. One may obtain a value for $ C_{n} $ by assuming that for large values of $\beta$, $ f(r) $ tends to the Yang-Mills solution.

\section{Thermodynamic quantities of the NYM black hole solutions}\label{thermo} 
According to the gauge/gravity duality, one can provide a relation between the strongly coupled gauge theories and the related weakly coupled string theories. From the holography viewpoint, the bulk string theory may inform the boundary gauge theory. Using AdS/CFT correspondence\cite{Malda, Witt} which maps the conformal field theory to the asymptotically AdS spacetime with a higher dimension, thermodynamic properties of a black hole may reveal the dual physical system properties. For instance, the horizon of a black hole in the asymptotically AdS spacetime can give information about the finite temperature of its dual field theory. \\
\indent In this part, we would like to investigate the
thermodynamic quantities of the NYM black
hole and then check the first law of thermodynamics. The Hawking temperature of the NYM black hole may be obtained from
\begin{eqnarray}\label{temp}
T_{+}&=&\frac{\kappa}{2\pi}=\frac{f^{'}(r_{+})}{4\pi}=\frac{k(n-3)}{4\pi r_{+}}-\frac{\Lambda}{2\pi(n-2)}r_{+}\nonumber\\
+&&\left\{
\begin{array}{ll}
$$+\frac{\beta^2 r_{+}}{\pi(n-2)}\bigg(1-\sqrt{1+\frac{(n-2)(n-3)e^2}{2\beta^2 r_{+}^4}}\bigg)$$,\quad\quad\quad\quad\quad\quad\,\,  \ {BI}\quad &  \\ \\
$$-\frac{\beta^2 r_{+}}{\pi(n-2)}\bigg[1-\mathrm{exp}\bigg(-\frac{(n-2)(n-3)e^2}{4\beta^2 r_{+}^4}\bigg)\bigg]$$,\quad\quad\quad \quad\quad\,  \ {EN}\quad &  \\ \\
$$-\frac{2\beta^2 r_{+}}{\pi(n-2)}\mathrm{ln}\bigg(1+\frac{(n-2)(n-3)e^2}{8\beta^2 r_{+}^4}\bigg)$$,\quad\quad\quad\quad\quad\quad\quad\quad\,\,  \ {LN}\quad &
\end{array}
\right.
\end{eqnarray}
where we have used $f(r_{+})=0$.
From the so-called area law (which states that the entropy is one-quarter of the event horizon area of the black hole \cite{Beken}), the entropy can be calculated as below
\begin{eqnarray}\label{entropy}
S=\frac{A}{4}=\frac{r_{+}^{n-2}}{4}\omega_{n-2},
\end{eqnarray}
where $\omega_{n-2}$ represents the volume of a $(n-2)$-dimensional unit sphere and a $(n-2)$-dimensional hypersurface with constant negative curvature for $ k=1 $ and $ k=-1 $ respectively.
To obtain the mass, we use the subtraction method of Brown and York \cite{York}. For this purpose, let us write the metric \eqref{metric} in the following form
\begin{eqnarray}
ds^2=\lambda_{ab}dx^{a}dx^{b}=-V(r) dt^2+\frac{dr^2}{V(r)}+r^2 d\Omega_{n-2}^{2},
\end{eqnarray} 
and choose a background metric with 
\begin{eqnarray}
V_{0}(r)=\left\{
\begin{array}{ll}
$$k-\frac{2\Lambda}{(n-1)(n-2)}r^2$$,\quad\quad\quad\quad\,\,\,  \ {n\neq5}\quad &  \\ \\
$$k-\frac{\Lambda}{6}r^2-\frac{2 e^2\mathrm{ln}(r/r_{0})}{r^2}$$\quad\quad\quad  \ {n=5.}\quad &
\end{array}
\right.
\end{eqnarray}
$V_{0}(r)$ is an arbitrary function that determines the zero of the energy to avoid the infinities of the mass. We again choose $r_{0}=1$. If we characterize $\sigma_{ab}$ as the metric of the spacelike surface $\Sigma$ in $\partial\mathcal{M}$, and $n^{a}$ and $\xi^{b}$ as the unit normal and the timelike killing vectors of this boundary, respectively, the mass of this black hole is calculated by
\begin{eqnarray}
M=\frac{1}{8\pi}\int_{\Sigma} d^{n-2}\sqrt{\sigma}\{(K_{ab}-K \lambda_{ab})-(K^{0}_{ab}-K^{0} \lambda^{0}_{ab})\}n^{a}\xi^{b},
\end{eqnarray} 
where $\sigma$ is the determinant of the metric $\sigma_{ab}$ and $K_{ab}^{0}$ is the extrinsic curvature tensor of the background metric. As we use the limit $r\rightarrow \infty$, the mass of the NYM black hole yields to
\begin{eqnarray}\label{Mass}
M=\frac{(n-2)\omega_{n-2}}{16\pi}m,
\end{eqnarray}
where $m$ is found from the equation $f(r_{+})=0$ in Eq. \eqref{fff}.\\
\indent The global Yang-Mills charge of the NYM black hole is obtained from the Gauss law
\begin{eqnarray}\label{charge}
Q=\frac{1}{4\pi}\int d^{n-2}x\sqrt{Tr(F_{\mu\nu}^{(a)}F^{(a)\mu\nu})}=\frac{\sqrt{(n-2)(n-3)}\,\omega_{n-2}}{4\pi}e.
\end{eqnarray}
 \indent We consider the mass $M$ in Eq. \eqref{Mass} as a function of the entropy \eqref{entropy} and the charge \eqref{charge}, so the first law of thermodynamics is specified by
\begin{eqnarray}
dM=TdS+\Phi dQ,
\end{eqnarray}
where $T=\big(\frac{\partial M}{\partial S}\big)_{Q}$ and $ \Phi$ is the gauge potential, $\Phi=\big(\frac{\partial M}{\partial Q}\big)_{S}$. Numerical calculations demonstrate that the evaluated $T$ is in agreement with $T_{+}$ in Eq. \eqref{temp}. So, the first law of the black hole thermodynamics is satisfied, if we obtain the gauge potential of the solutions with $n\neq 4z+1$ as below
\begin{eqnarray}\label{electriccharge}
\Phi=\bigg(\frac{\partial M}{\partial Q}\bigg)_{S}=\left\{
\begin{array}{ll}
$$-\frac{2\pi Q (n-2)(n-3)r_{+}^{n-5}}{(n-5)}{}_2F_{1}\big(\big[\frac{1}{2},\frac{5-n}{4}\big]\,,\big[\frac{9-n}{4}\big]\,,-8\xi_{+}\big)$$,\quad\quad\quad\quad\quad\quad\quad\quad\quad\quad\quad \quad\quad\ {BI}\quad &  \\ \\
$$-\frac{2\pi Q (n-2)(n-3)r_{+}^{n-5}}{(n-5)}{}_2F_{1}\big(\big[\frac{5-n}{4}\big]\,,\big[\frac{9-n}{4}\big]\,,-4\xi_{+}\big)$$,\quad\quad\quad\quad\quad\quad\quad \quad\quad\quad\quad\quad\quad\quad  \ {EN}\quad &  \\ \\
$$-\frac{2\pi Q(n-2)(n-3)r_{+}^{n-5}}{(n-1)(1+2\xi_{+})}-\frac{8\pi Q (n-2)(n-3)r_{+}^{n-5}}{(n-1)(n-5)}{}_2F_{1}\big(\big[1,\frac{5-n}{4}\big]\,,\big[\frac{9-n}{4}\big]\,,-2\xi_{+}\big)+\\ \frac{16\pi Q\xi_{+}(n-2)(n-3)r_{+}^{n-5}}{(n-1)(n-9)}{}_2F_{1}\big(\big[2,\frac{9-n}{4}\big]\,,\big[\frac{13-n}{4}\big]\,,-2\xi_{+}\big)$$.\quad\quad\quad\quad\quad\quad\quad\quad\quad\quad\quad\quad\,\,  \ {LN}\quad &
\end{array}
\right.
\end{eqnarray}
where $\xi_{+}=\frac{(n-2)(n-3)\pi^2 Q^2}{\beta^2 r_{+}^4}$. The gauge potential of the solutions with $n=4z+1$ can be derived exactly for each dimension, however it does not have a general form.
\section{Thermal stability of the NYM black hole solutions}\label{stability} 
\begin{figure}
\centering
\subfigure[\,$n=4$]{\includegraphics[scale=0.27]{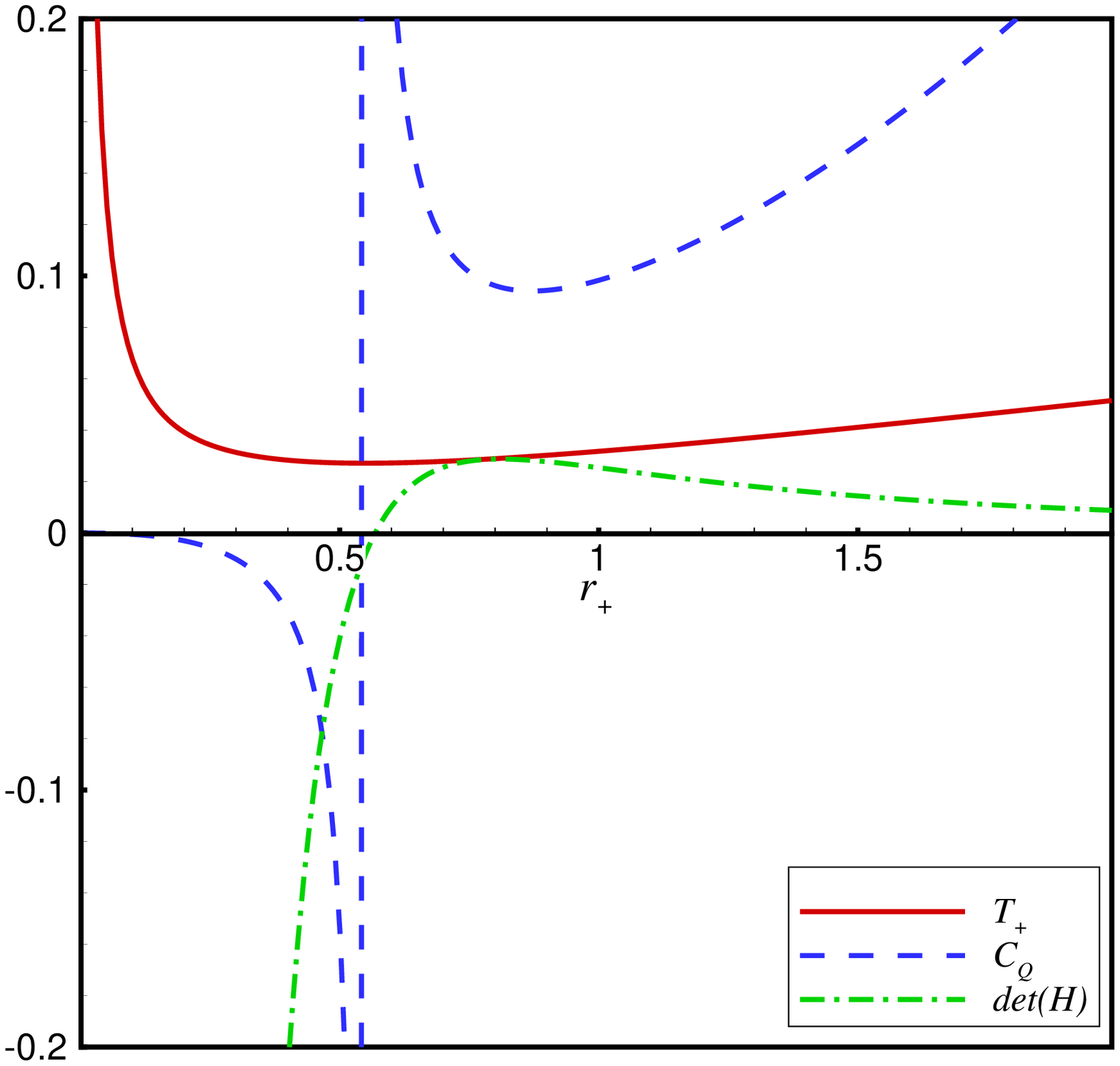}\label{Fig3a}}\hspace*{.2cm}
\subfigure[\,$n=5$]{\includegraphics[scale=0.27]{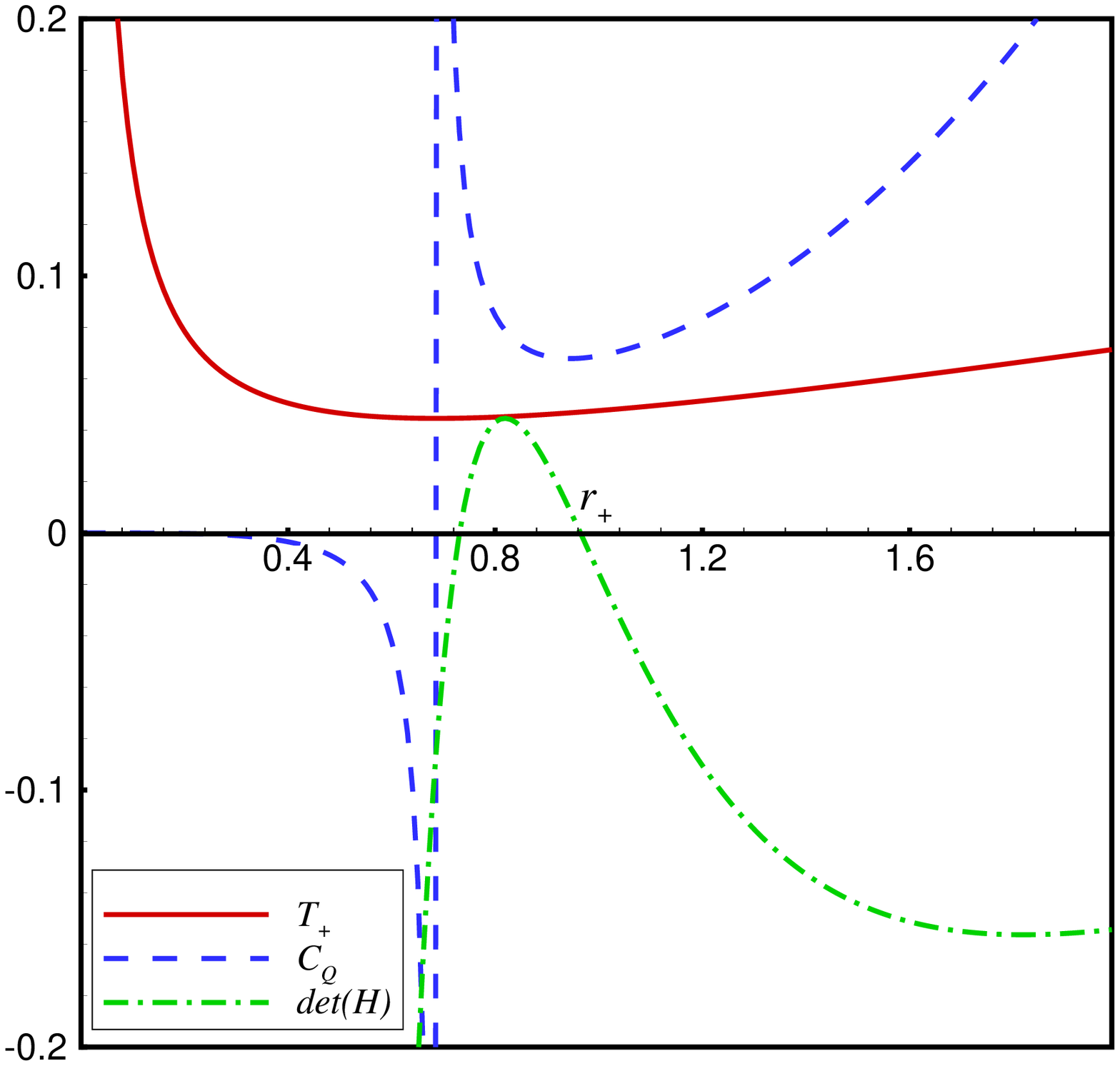}\label{Fig3b}}\hspace*{.2cm}
\subfigure[\,$n=6$]{\includegraphics[scale=0.27]{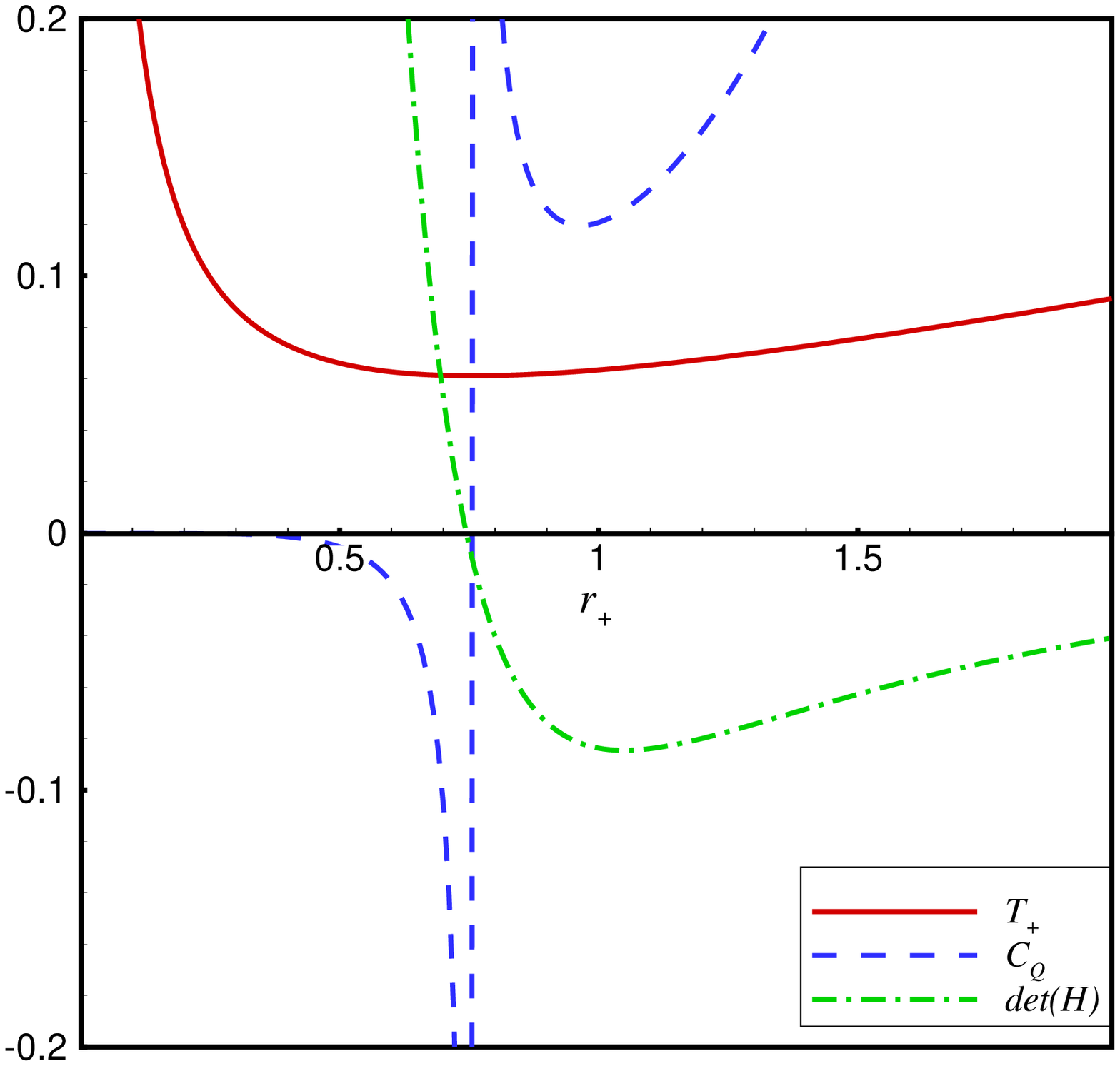}\label{Fig3c}}\caption{$T_{+}$, $C_{Q}$ and $det(H)$ versus $r_{+}$ for different dimensions $n$ with $\Lambda=-(n-1)(n-2)/2l^2$, $l=1$, $k=1$, $e=0.1$ and $\beta=1$ for BIYM theory.}\label{Fig3}
\end{figure}

\begin{figure}
\centering
\subfigure[\,$n=4$]{\includegraphics[scale=0.27]{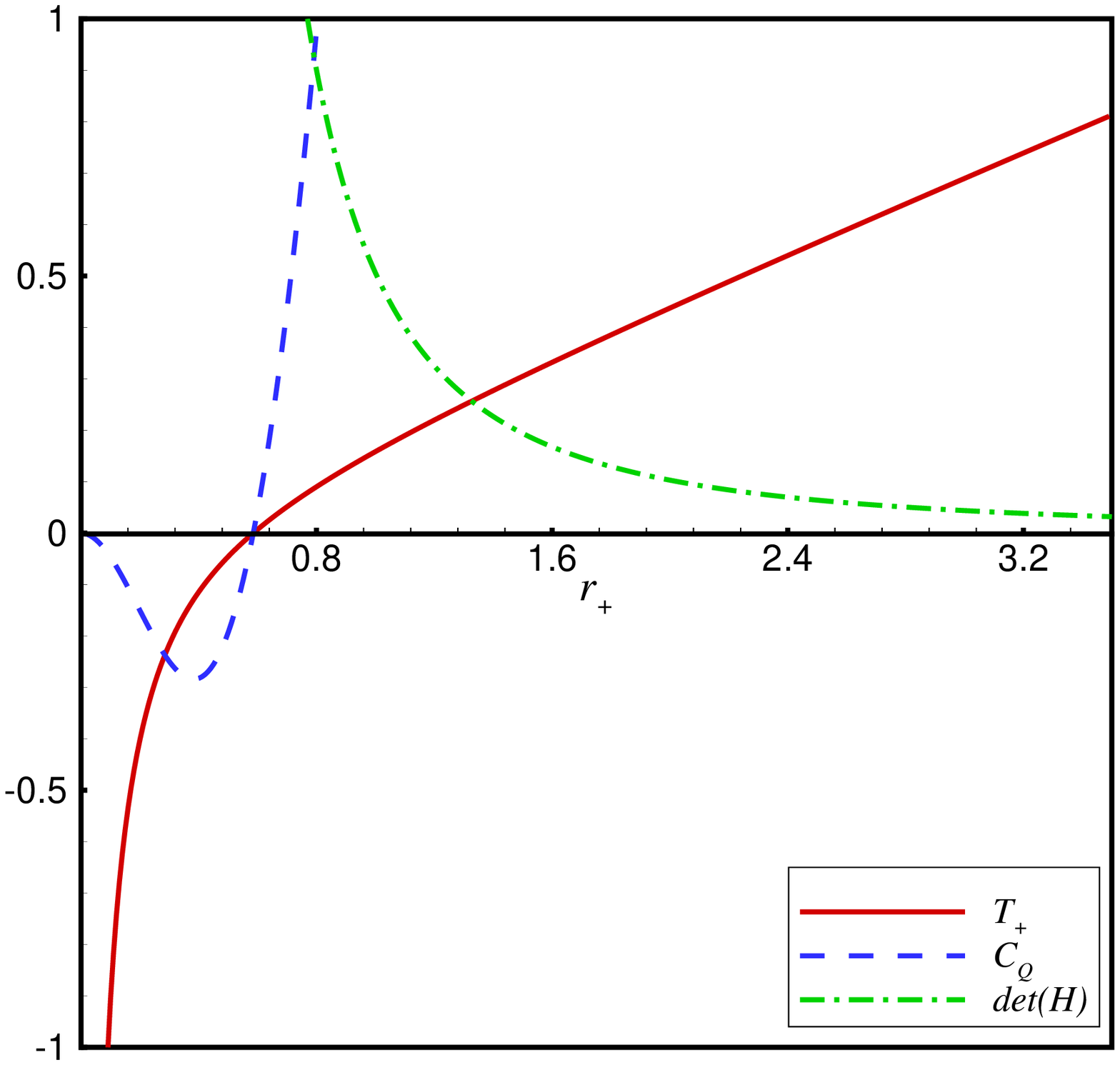}\label{Fig4a}}\hspace*{.2cm}
\subfigure[\,$n=5$]{\includegraphics[scale=0.27]{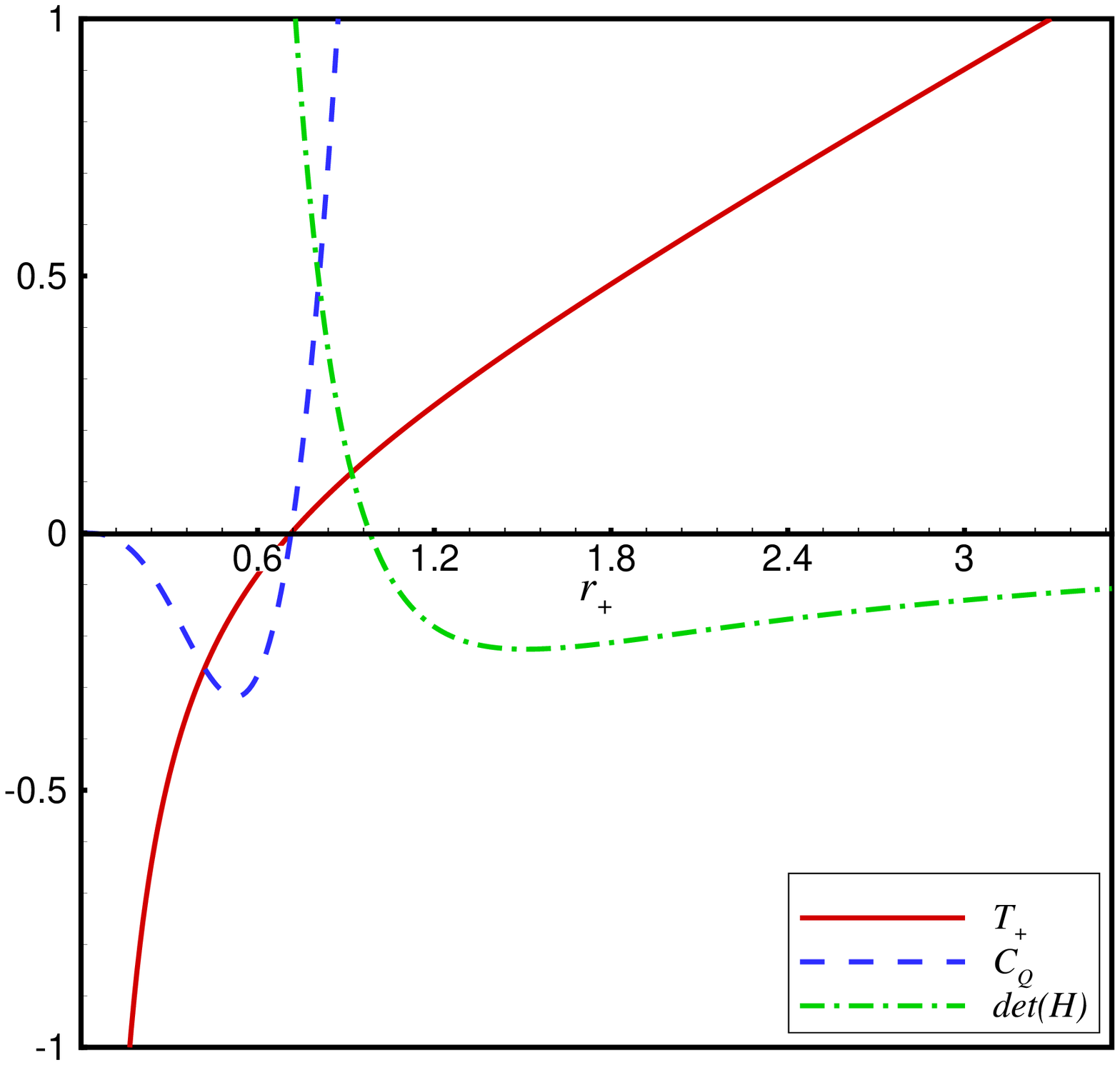}\label{Fig4b}}\hspace*{.2cm}
\subfigure[\,$n=6$]{\includegraphics[scale=0.27]{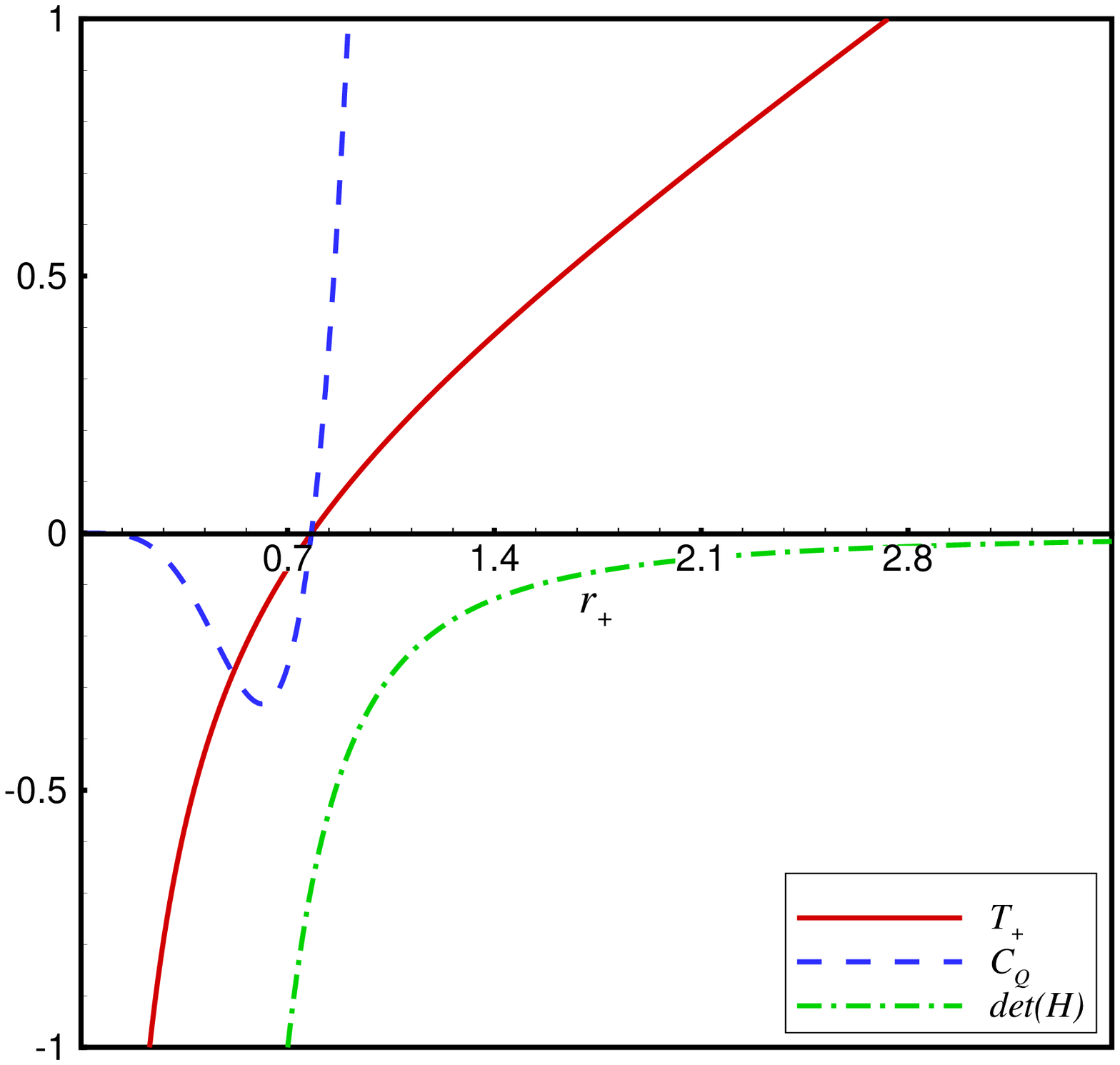}\label{Fig4c}}\caption{$T_{+}$, $C_{Q}$ and $det(H)$ versus $r_{+}$ for different dimensions $n$ with $\Lambda=-(n-1)(n-2)/2l^2$, $l=1$, $k=-1$, $e=0.1$ and $\beta=1$ for ENYM theory.}\label{Fig4}
\end{figure}

\begin{figure}
\centering
\subfigure[\,$n=4$]{\includegraphics[scale=0.27]{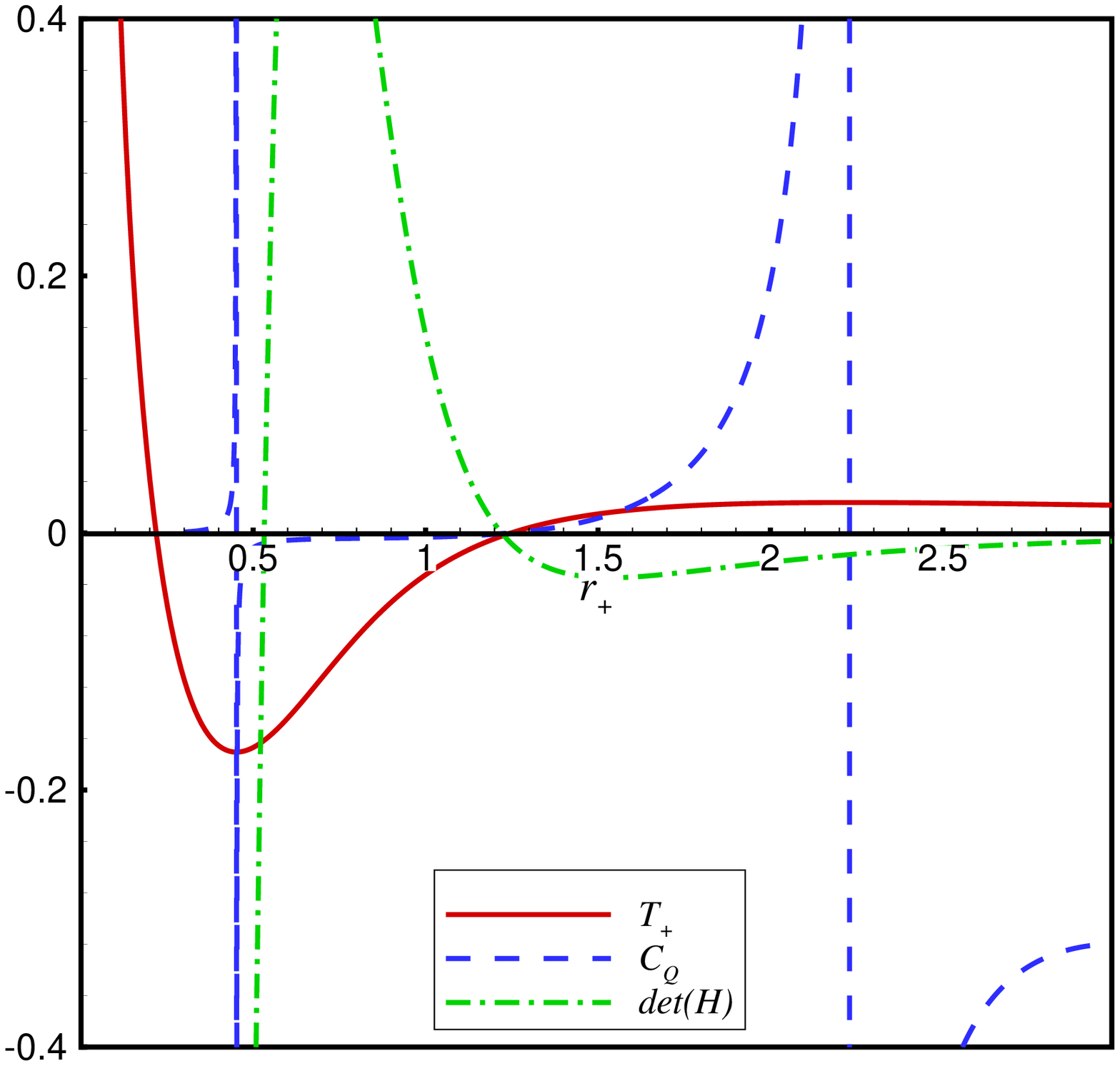}\label{Fig5a}}\hspace*{.2cm}
\subfigure[\,$n=5$]{\includegraphics[scale=0.27]{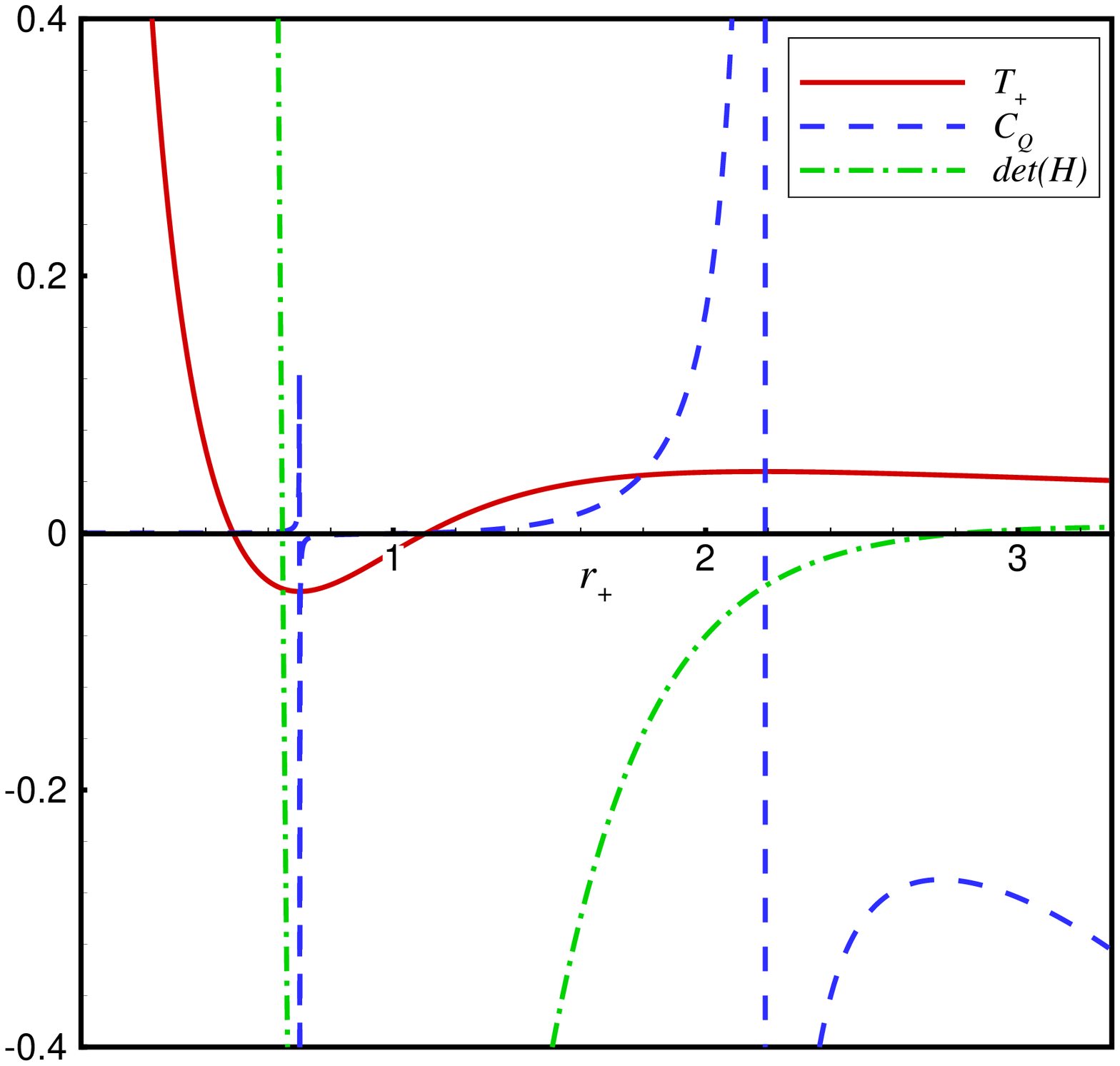}\label{Fig5b}}\hspace*{.2cm}
\subfigure[\,$n=6$]{\includegraphics[scale=0.27]{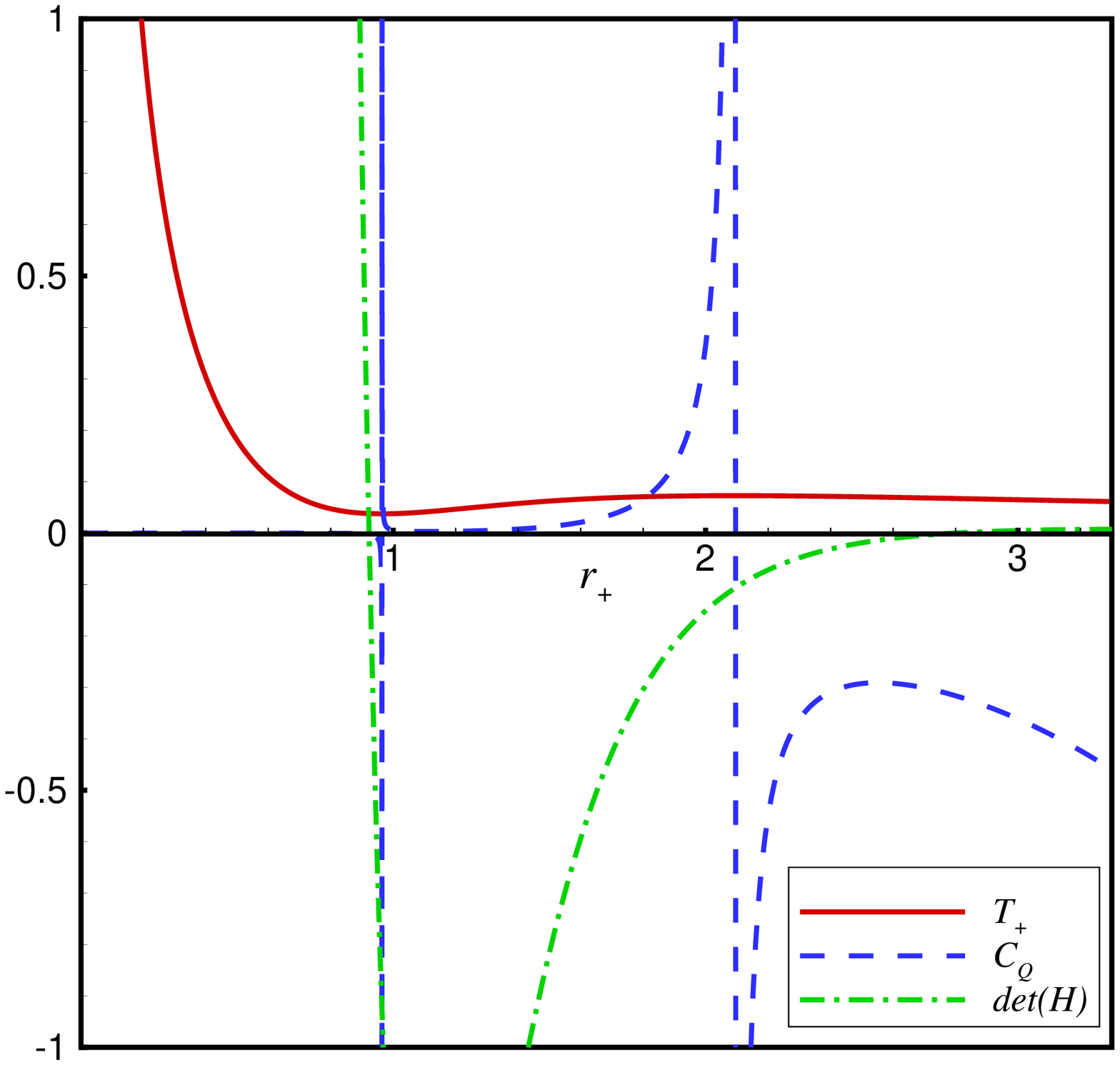}\label{Fig5c}}\caption{$T_{+}$, $C_{Q}$ and $det(H)$ versus $r_{+}$ for different dimensions $n$ with $\Lambda=0$, $k=1$, $e=1.3$ and $\beta=1$ for LNYM theory.}\label{Fig5}
\end{figure}

\begin{figure}
\centering
\includegraphics[scale=0.5]{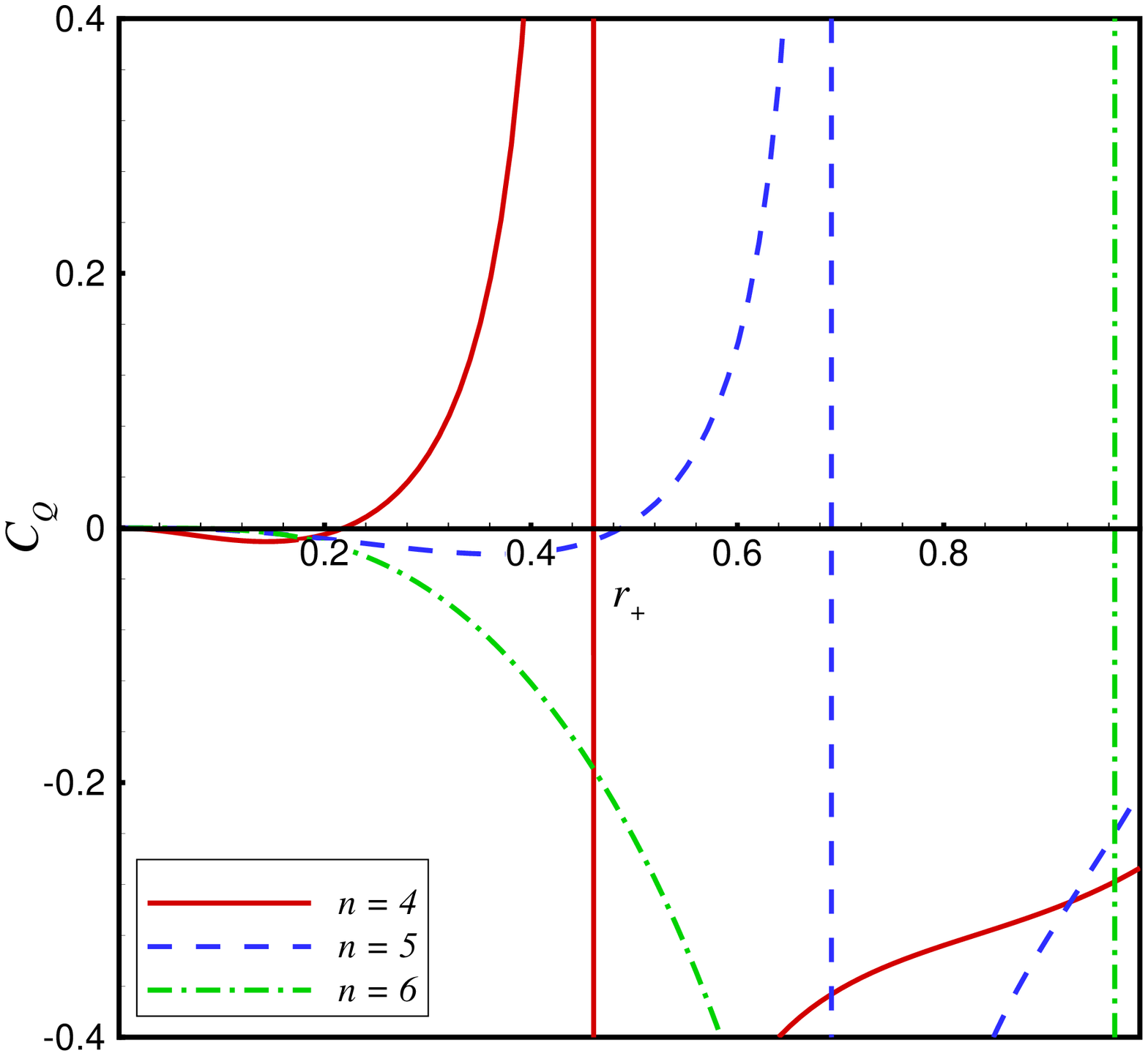}
\caption{\small{$C_{Q}$ versus $r_{+}$ ($0\le r_{+}\le 1$) for different dimensions $n$ with $\Lambda=0$, $k=1$, $e=1.3$ and $\beta=1$ for LNYM theory.} \label{Fig6}}
\end{figure}

The thermal stability of a black hole is determined if one analyzes the behavior of the entropy $S$ or the energy $M$ with respect to the small variations of the thermodynamic coordinates around the equilibrium. In this section, we aim to consider $S$ and $Q$ as a set of thermodynamic variables and probe the NYM black hole thermal stability in the canonical and grand canonical ensembles. In the canonical ensemble, the parameter charge $Q$ is fixed and so the black hole is thermally stable if the heat capacity
\begin{eqnarray}\label{cano}
C_{Q}=T_{+}\big(\frac{\partial S}{\partial T_{+}}\big)_{Q}=T_{+}\bigg(\frac{\partial^2 M}{\partial S^2}\bigg)^{-1}_{Q},
\end{eqnarray}
is positive. It should be noted that in order to have physical solutions, the temperature should be also positive. So, a physically stable NYM black hole in the canonical ensemble is obtained, if the conditions $T_{+}>0$ and $C_{Q}>0$ are satisfied. In the grand canonical ensemble, both parameters $S$ and $Q$ are variables and so the positive value of the Hessian matrix determinant may lead to stable solutions. The Hessian matrix is defined as 
\begin{eqnarray}\label{Gcano}
H=\left[
\begin{array}{ccc}
\Big(\frac{\partial ^2 M}{\partial S^2}\Big)_{Q} & \Big(\frac{\partial ^2 M}{\partial S\partial Q}\Big)\\
\Big(\frac{\partial ^2 M}{\partial Q\partial S}\Big) & \Big(\frac{\partial ^2 M}{\partial Q^2}\Big)_{S}
\end{array} \right],
\end{eqnarray}
where we abbreviate the related determinant to $det(H)$. In this ensemble, the two conditions $\big(\frac{\partial^2 M}{\partial S^2}\big)_{Q}>0$ and $\big(\frac{\partial^2 M}{\partial Q^2}\big)_{S}>0$ should be satisfied. If all three quantities, temperature $T_{+}$, heat capacity $C_{Q}$ and $det(H)$ are positive, then these two conditions are established spontaneously from Eqs. \eqref{cano} and \eqref{Gcano}.\\
\indent To find a physical stable region for the NYM black hole in both canonical and grand canonical ensembles, we have plotted the temperature $T_{+}$, heat capacity $C_{Q}$ and Hessian-matrix determinant $det(H)$ versus $r_{+}$ in Figs. (\ref{Fig3}-\ref{Fig6}). As the third term in Eq. \eqref{temp} is negative for the three BIYM, ENYM, and LNYM cases, a positive value for the temperature depends on the values of the first two terms. To reduce the effect of the third term in Eq. \eqref{temp}, we choose a small value for the parameter $Q$. We can also find from Eq. \eqref{temp} that the temperature is positive just for the AdS ($\Lambda<0$) solutions with $k=\pm 1$ , and also for dS($\Lambda>0$) and flat ($\Lambda=0$) solutions with $k=1$. We have probed the solutions with these features in Figs. (\ref{Fig3}-\ref{Fig6}). We refuse to investigate the thermal stability of the dS solutions since our results have shown that it is not possible to obtain a positive region for both quantities $T_{+}$ and $C_{Q}$. \\
For an economic reason, we have studied just the stability of the AdS-BIYM solutions with $k=1$ in Fig. (\ref{Fig3}), the AdS-ENYM solutions with $k=-1$ in Fig. (\ref{Fig4}) and flat LNYM solutions with $k=1$ in Figs. (\ref{Fig5}) and (\ref{Fig6}).\\
\indent In Fig. (\ref{Fig3}), the temperature is positive for all values of $r_{+}$ in dimensions $n=4,5,6$. So, the thermal stability of these solutions in the canonical ensemble depends only on the heat capacity value. There is a $r_{+\mathrm{min1}}$ which the heat capacity is positive for $r_{+}>r_{+\mathrm{min1}}$ and it increases as the dimension $n$ increases. To obtain a physically stable region in the grand canonical ensemble, all the quantities $T_{+}$, $C_{Q}$ and $det(H)$ should be positive. We can conclude from Fig. (\ref{Fig3}) that the stable region becomes smaller as the dimension $n$ increases. For example, the four-dimensional BIYM solutions are stable for $r_{+}>r_{+\mathrm{min1}}$, while there is no stable regions for $n=6$. For $n=5$, there is just a small stable region for the BIYM black hole. Obviously, the obtained stable regions are different in these two ensembles, which is expected. However, one may choose the cosmological constant as a thermodynamics variable. It is argued in \cite{seyedh} that considering cosmological constant as a variable may lead to the same stable regions for the two ensembles.  \\
\indent For the AdS-ENYM black hole in Fig. (\ref{Fig4}), there is a $r_{+\mathrm{min2}}$ for each dimensions $n=4,5,6$, which both $T_{+}$ and $C_{Q}$ are positive for $r_{+}>r_{+\mathrm{min2}}$. By increasing the dimension $n$, the positive value of $det(H)$ decreases. The four-dimensional AdS-ENYM solutions with $k=-1$ are thermally stable for $r_{+}>r_{+\mathrm{min2}}$ in the grand canonical ensemble, however, there is no thermal stability for $n=6$.\\
\indent In Fig. (\ref{Fig5}), we have probed the stability of the flat LNYM solutions with $k=1$. As the heat capacity behavior is not clear in $n=4,5,6$ for the range of $0\le r_{+}\le 1$, so we have magnified it in Fig. (\ref{Fig6}).  Figs. (\ref{Fig5}) and (\ref{Fig6}) show that there are two values $r_{+\mathrm{min3}}$ and $r_{+\mathrm{max}}$ which both $T_{+}$ and $C_{Q}$ are positive for $r_{+\mathrm{min3}}\le r_{+}\le r_{+\mathrm{max}}$. As the dimension $n$ increases, the values of $r_{+\mathrm{min3}}$ and $r_{+\mathrm{max}}$ decrease. As the quantity $det(H)$ is negative for the range $r_{+\mathrm{min3}}\le r_{+}\le r_{+\mathrm{max}}$, so we cannot find a stable region for the flat solutions in the grand canonical ensemble. 
\section{critical behavior of the NYM black hole solutions}\label{P-V}
In this section, we would like to study the critical behavior and phase transitions of the NYM black holes in the extended phase space. One can enlarge the thermodynamic
phase space and consider the cosmological constant as a thermodynamic pressure \cite{Tras, Dolan1, Dolan2}. Hawking and Page were the first who showed a phase transition for the Schwarzschild AdS black hole \cite{Page1}. Recently, many studies about the critical behavior and phase transition of black holes have been done \cite{Kubiz10, Mirza1, Kamrani, Mirza2}. The critical behavior of the BI Maxwell (BIM) black hole in the AdS spacetime has been also investigated \cite{Kubiznak, Fernando, Fernando1}. In Ref.\cite{Dayyani}, the critical behavior of the BI-dilaton black holes has been discussed. In the following, we first obtain a Smarr relation, then we get to an equation of state and a Gibbs energy to check out the phase transition of the NYM black hole. We also obtain the critical exponents of this black hole and compare them with the Van der Waals fluid.     
\subsection{Smarr relation}
To obtain a Smarr relation, we should investigate the thermodynamics of the black hole in the extended phase space. For the NYM black hole, we consider the quantities $S$ and $Q$, the dimensionful parameters $\Lambda$ and $\beta$, and their conjugates as thermodynamic variables. In this way, we can write the first law of thermodynamics in the extended phase space
\begin{eqnarray}
dM=TdS+\Phi dQ+VdP+B d\beta,
\end{eqnarray}  
where the pressure $P$ is defined as 
\begin{eqnarray}
\label{pressure}
P=-\frac{\Lambda}{8\pi}.
\end{eqnarray}
\indent If we consider the specific volume $v=\frac{4r_{+}}{n-2}$ and use Eq. \eqref{Mass}, then the conjugate quantity of $P$ is
\begin{eqnarray}\label{volume}
V=\frac{\omega_{n-2}}{n-1}r_{+}^{n-1},
\end{eqnarray}
and the related conjugate of $\beta$ in the $n\neq 4z+1$ case is defined as 
\begin{eqnarray}\label{BB}
B&=&\frac{\partial M}{\partial \beta}\nonumber\\
&&=\left\{
\begin{array}{ll}
$$\frac{e^2(n-2)(n-3)r^{n-5}}{8\pi\beta(n-5)}{}_2F_{1}\big(\big[\frac{1}{2},\frac{5-n}{4}\big]\,,\big[\frac{9-n}{4}\big]\,,-\frac{\eta_{+}}{2}\big)-\frac{\beta r^{n-1}}{2\pi(n-1)}\big({}_2F_{1}\big(\big[-\frac{1}{2},\frac{1-n}{4}\big]\,,\big[\frac{5-n}{4}\big]\,,-\frac{\eta_{+}}{2}\big)-1\big)$$,\quad\quad\quad  \ {BI}\quad &  \\ \\
$$\frac{e^2(n-2)(n-3)r^{n-5}}{8\pi\beta(n-5)} {}_2F_{1}\big(\big[\frac{5-n}{4}\big]\,,\big[\frac{9-n}{4}\big]\,,-\frac{\eta_{+}}{4}\big)+\frac{\beta r^{n-1}}{2\pi(n-1)}\big({}_2F_{1}\big(\big[\frac{1-n}{4}\big]\,,\big[\frac{5-n}{4}\big]\,,-\frac{\eta_{+}}{4}\big)-1\big)$$,\quad\quad\quad \quad\quad \quad\,\, \ {EN}\quad &  \\ \\
$$\frac{(n-2)(n-3)e^2 r^{n-5}}{8\pi\beta(n-1)}(1+\frac{\eta_{+}}{8})^{-1}-\frac{\beta r ^{n-1}}{\pi(n-1)}\mathrm{ln} (1+\frac{\eta_{+}}{8})-\frac{\eta_{+}(n-2)(n-3) e^2 r^{n-5}}{16\pi(n-1)(n-9)\beta}{}_2F_{1}\big(\big[2,\frac{9-n}{4}\big]\,,\big[\frac{13-n}{4}\big]\,,-\frac{\eta_{+}}{8}\big)$$.\,\, \ {LN}\quad &
\end{array}
\right.
\end{eqnarray}
\indent By using Eqs. \eqref{temp}, \eqref{entropy}, \eqref{Mass}, \eqref{charge}, \eqref{electriccharge}, \eqref{pressure}, \eqref{volume} and \eqref{BB}, the Smarr relation of the NYM black hole for $n\neq4z+1$ can be derived as 
\begin{eqnarray}\label{sssmar}
M=\frac{1}{n-3}[\Phi Q-\beta B-2PV+(n-2)TS],  \,\,\,\,\,\,\,\, \mathrm{for}\,\,\, n\neq 4z+1.
\end{eqnarray}
\indent It should be noted that the Smarr relation is not satisfied for the case $n=4z+1$. This is not unexpected as there are some other black hole solutions in the context of nonlinear electrodynamics for which the Smarr relation is not satisfied\cite{balartfer,breton,rasheed}. It was argued that the reason is that the trace of energy momentum tensor is not zero. The trace of energy momentum tensor is not zero for the NYM black holes. However, we have Smarr relation for $n\neq4z+1$. We guess it may originate from some properties of hypergeometric functions which have no explicit form for the case $n\neq4z+1$. We hope to investigate this issue in the future and find a physical reason for the Smarr relation not being satisfied.
\subsection{Equation of state}
\begin{figure}
\centering
\includegraphics[scale=0.5]{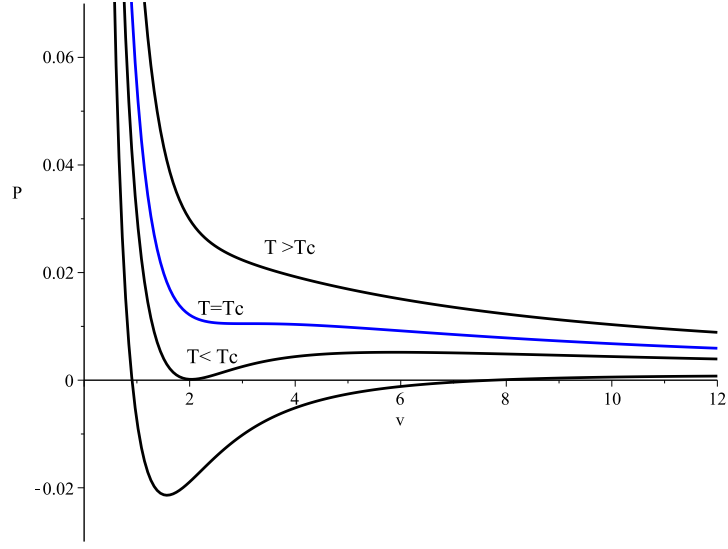}
\caption{\small{P-v diagram of BIYM theory for different values of temperature $T$ with $e=1$, $\beta=1 $, $n=5$ and $k=1$.} \label{Fig7}}
\end{figure}

\begin{figure}
\centering
\includegraphics[scale=0.5]{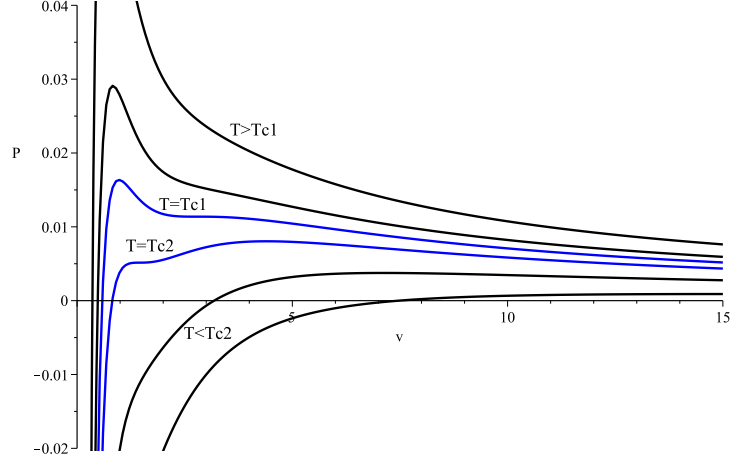}
\caption{\small{P-v diagram of BIYM theory for different values of temperature $T$ with $ \beta \in (\beta_{0},\beta_{1})$ and $ n=5 $. We have set $ e=1 $ and $\beta=0.7$ for which $ T_{c1}\approx 0.091626 $ and $ T_{c2}\approx 0.079242 $.} \label{Fig8}}
\end{figure}

\begin{figure}
\centering
\includegraphics[scale=0.5]{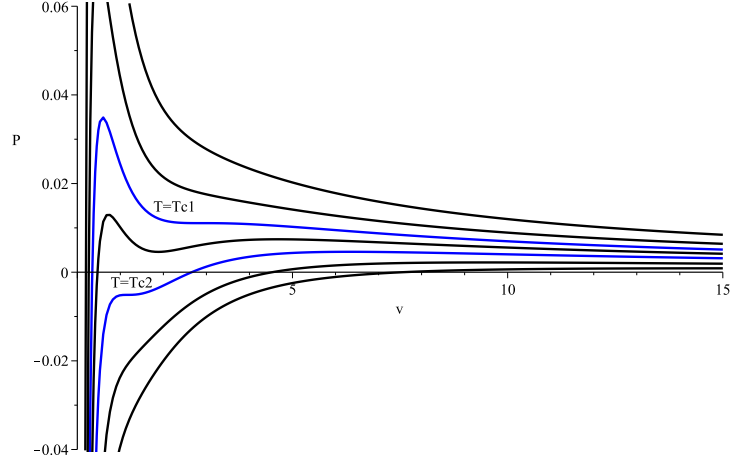}
\caption{\small{P-v diagram of BIYM theory for $ \beta \in (\beta_{1},\beta_{2})$ and $ n=5 $. We have set $ e=1 $ and $\beta=0.77$ for which $ T_{c1}\approx 0.090607 $ and $ T_{c2}\approx 0.061157 $.} \label{Fig9}}
\end{figure}

\begin{figure}
\centering
\subfigure[\,Critical temprature]{\includegraphics[scale=0.27]{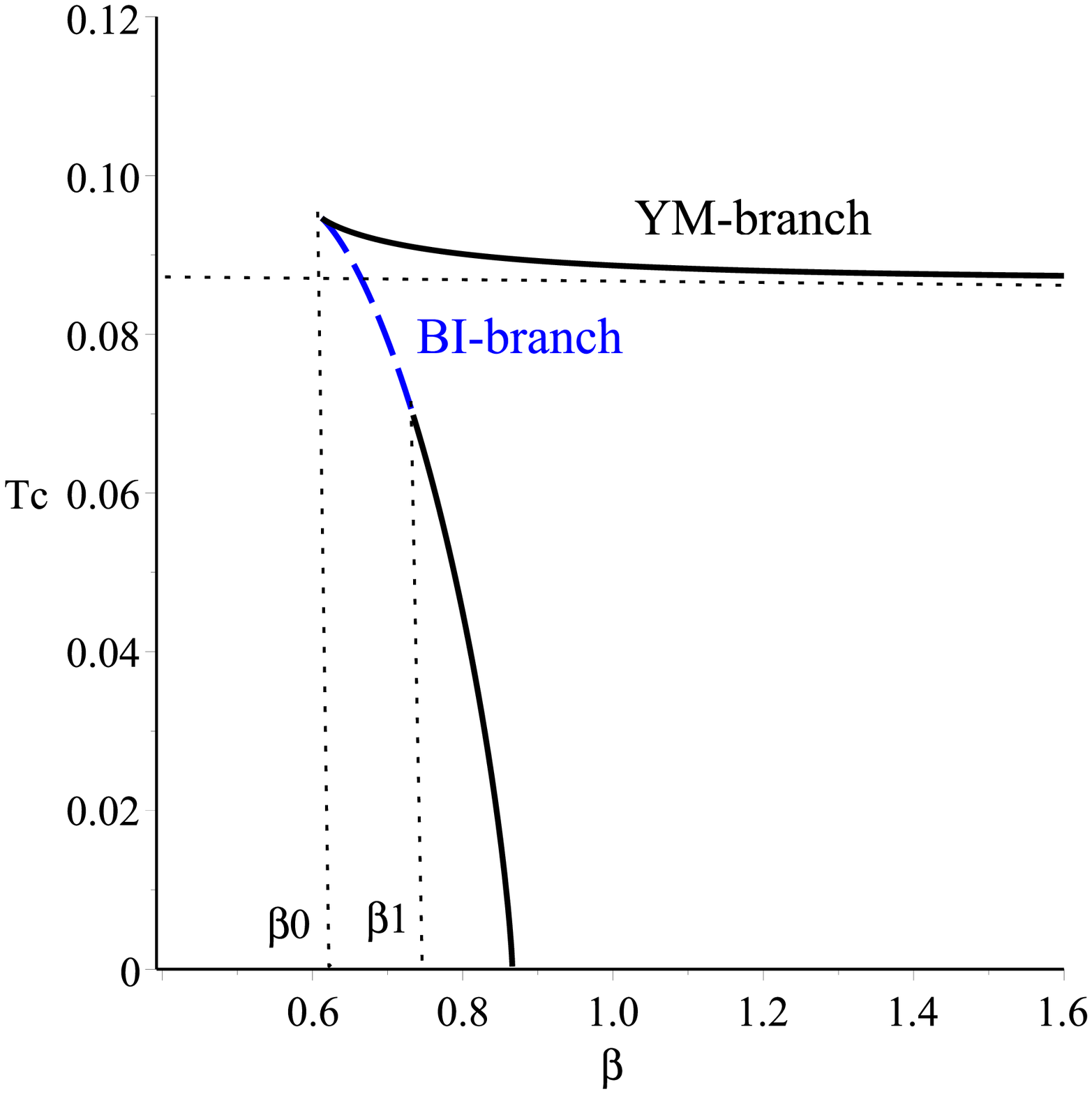}\label{Fig10a}}\hspace*{.2cm}
\subfigure[\,Critical volume]{\includegraphics[scale=0.27]{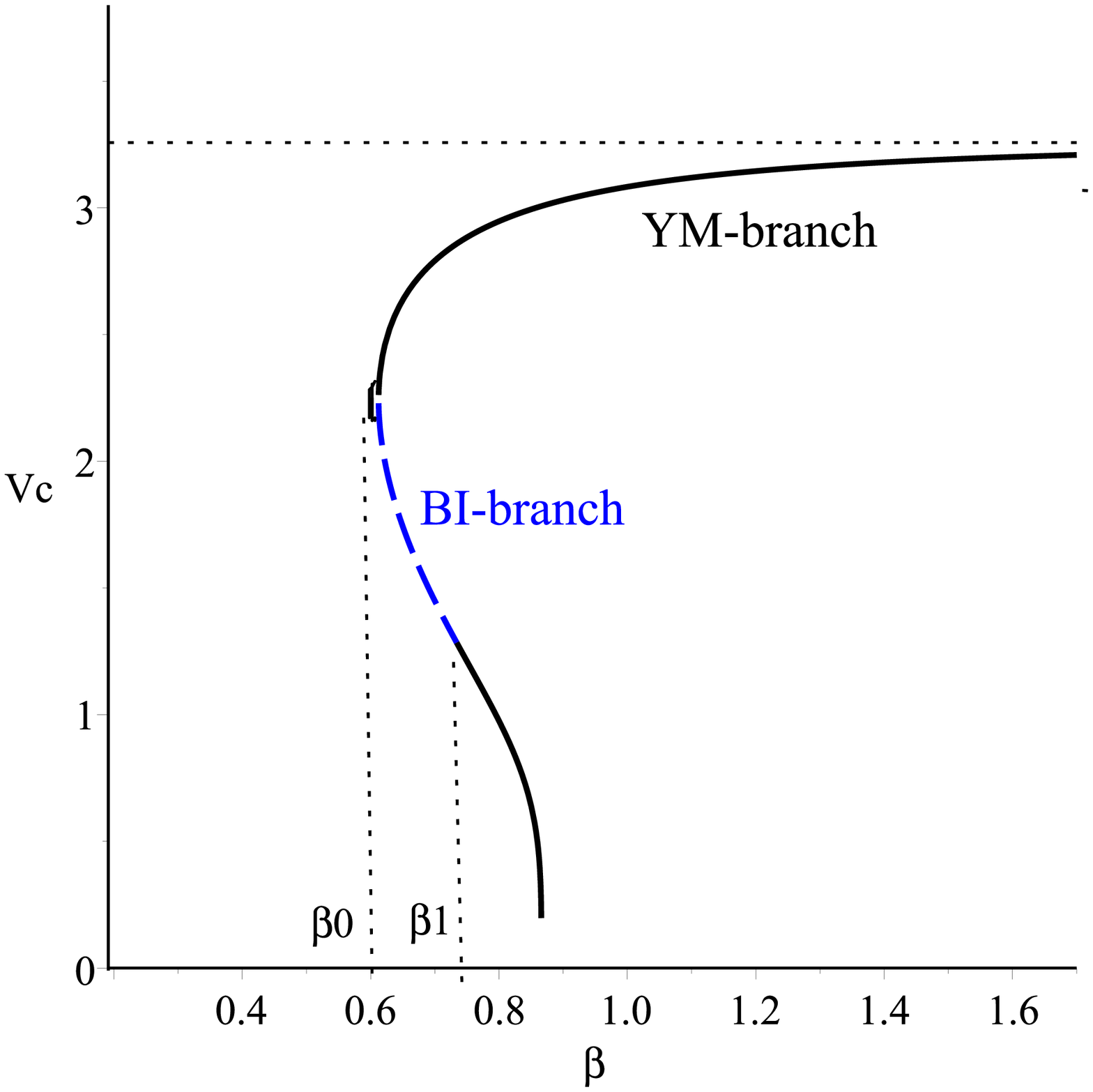}\label{Fig10b}}\hspace*{.2cm}	\subfigure[\,Critical pressure]{\includegraphics[scale=0.27]{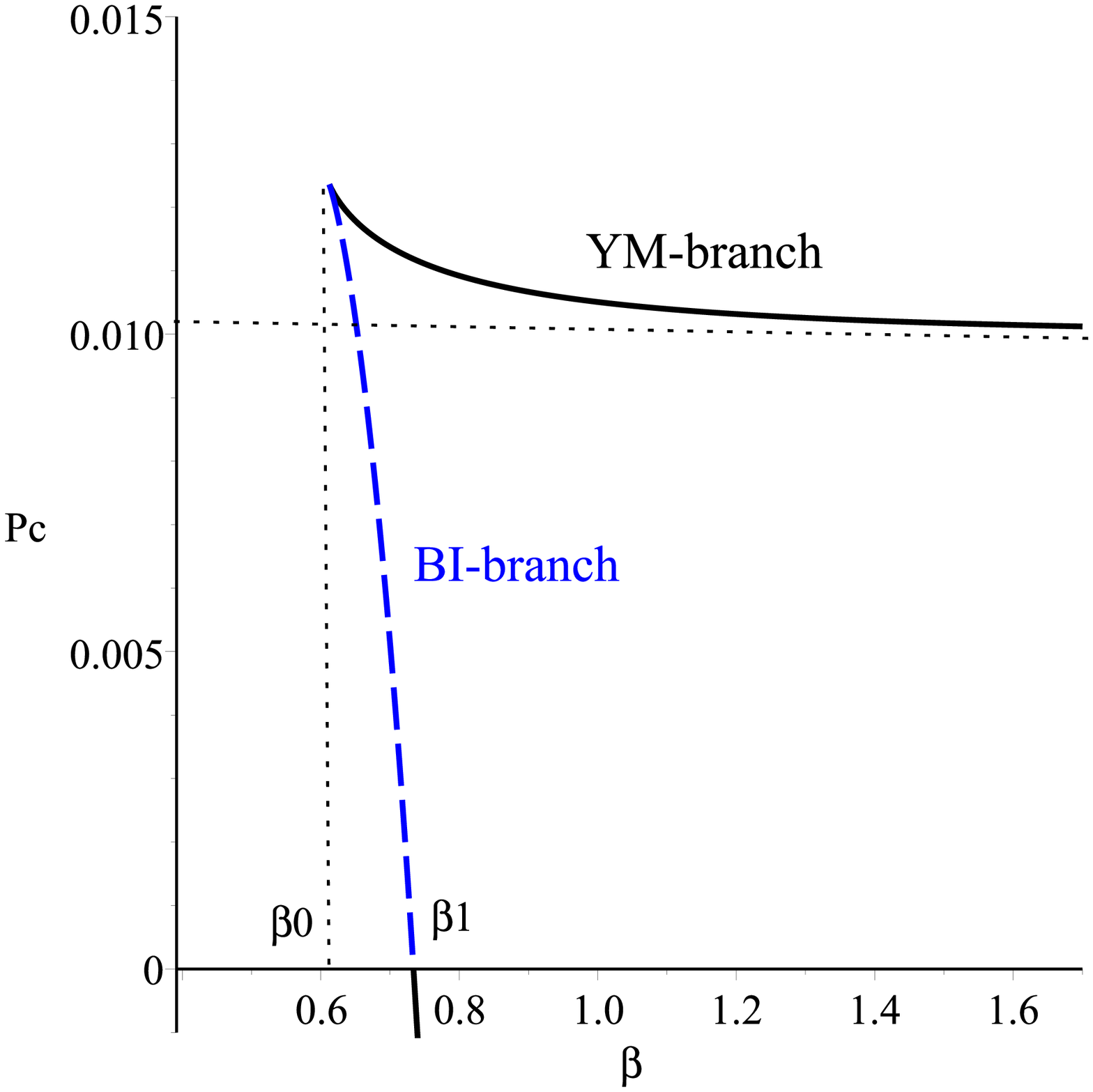}\label{Fig10c}}\caption{critical values $T_{c}$, $v_{c}$ and $P_{c}$ versus $\beta $ for $ n=5 $ and $ e=1 $. The black and blue solid lines indicate the YM and BI-branches respectively. The BI-branch corresponds to $ \beta\in(\beta_{0},\beta_{1}) $. The dashed horizontal lines are the critical values for the YM-AdS black hole for which $ T_{c,YM}=\sqrt{6}/9\pi e $, $v_{c,YM}=4 \sqrt{6} e/3 $ and $P_{c}=1/32 \pi e^2$. They can be obtained from the limit $\beta\rightarrow\infty$.}\label{Fig10}
\end{figure}

\begin{figure}
\centering
\includegraphics[scale=0.3]{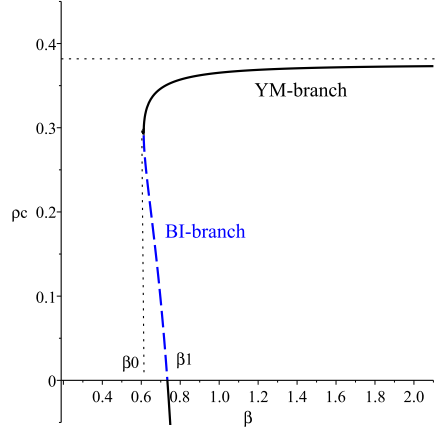}
\caption{\small{The critical ratio $\rho_{c}$ with respect to $\beta$ for $ n=5 $ and $ e=1 $. For the limit $ \beta\rightarrow\infty $, the critical ratio asymptotes to the YM-AdS one with $\rho_{c,YM}=\frac{3}{8} $. It has a unique value for the YM-AdS black hole in each dimension.} \label{Fig11}}
\end{figure}

\begin{figure}
\centering
\subfigure[\,$ \beta \in (\beta_{0},\beta_{2})$ and  $\beta=0.9 $]{\includegraphics[scale=0.3]{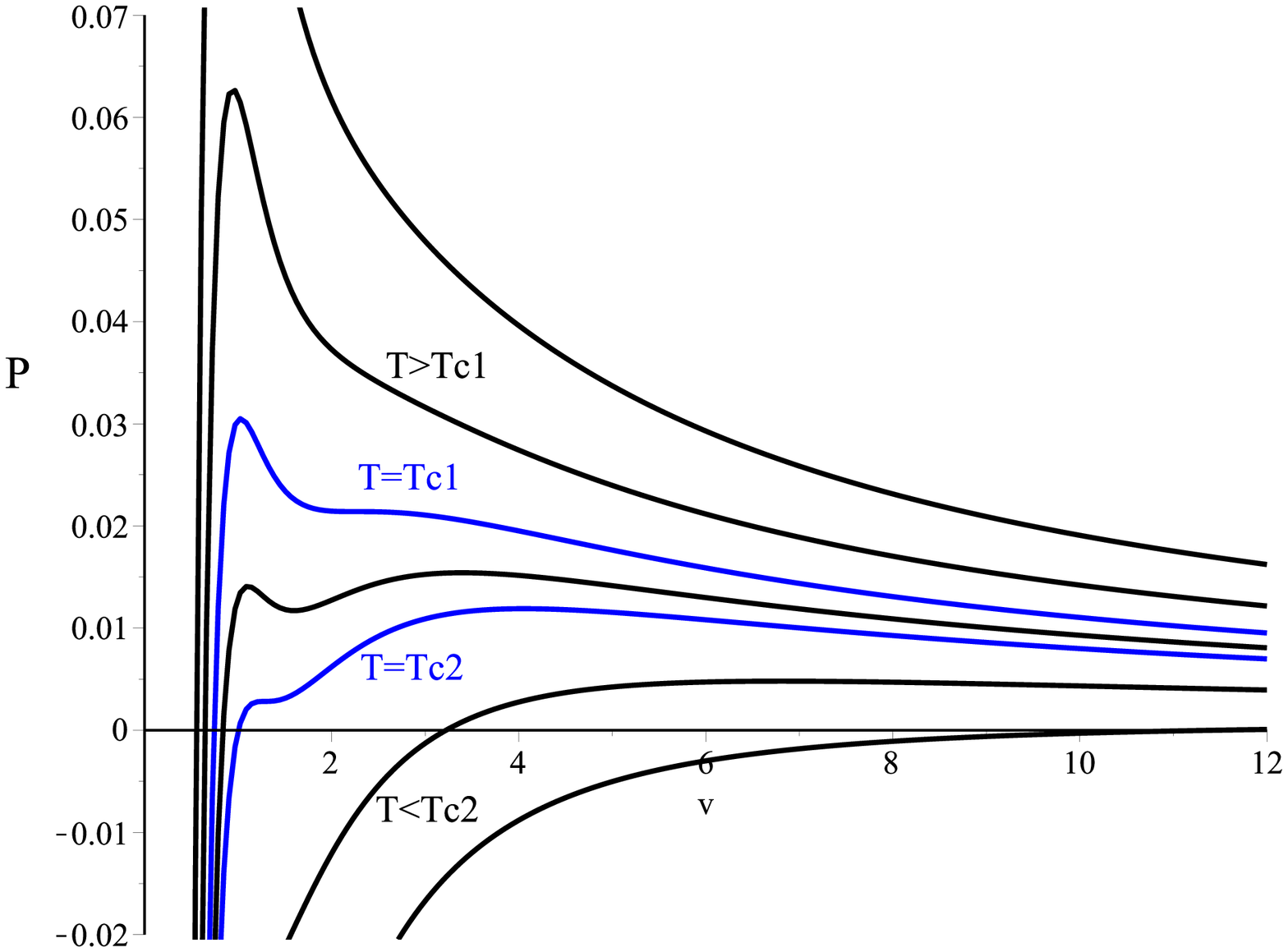}\label{Fig12a}}\hspace*{.5cm}
\subfigure[\, $\beta > \beta_{2}$ and $ \beta=10 $]{\includegraphics[scale=0.27]{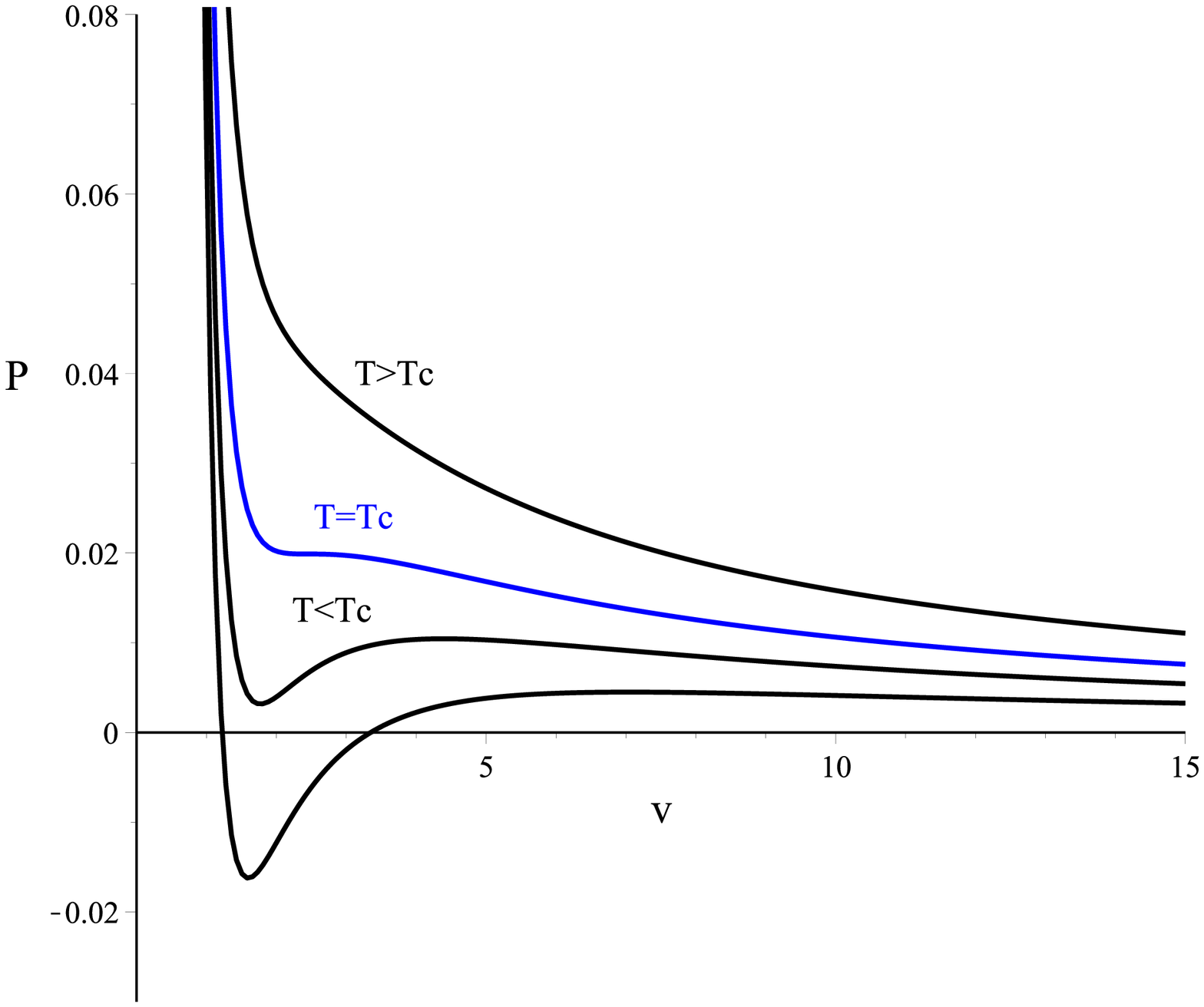}\label{Fig12b}}\caption{ P-v diagram of LNYM theory in $ n=6 $. We have set $ e=1 $ and $k=1$ .}\label{Fig12}
\end{figure}

To compare the critical behavior of the NYM black hole with that of the Van der Waals fluid, we first obtain the equation of state $P(T,v,\beta)\equiv P$, using equation \eqref{temp}. The critical points may be determined by using the following conditions  
\begin{eqnarray}\label{critical}
\frac{\partial P}{\partial v}=0\,\,\,,\,\,\,\frac{\partial^2 P}{\partial v^2}=0.
\end{eqnarray}
\indent We denote the volume, temperature and pressure of the critical points by $v_{c}$, $T_{c}$ and $P_{c}$, respectively. In the following, we will discuss the critical behavior of the NYM black hole for three types BIYM, ENYM, and LNYM, separately:\\
\textbf{Critical behavior of the BIYM solutions}\\
\indent By substituting the relation \eqref{pressure} in Eq. \eqref{temp}, one can find the equation of state for the BIYM black hole,
\begin{eqnarray}\label{p1}
P=\frac{T}{v}-k\frac{n-3}{\pi(n-2)v^2}-\frac{\beta^2}{4\pi}\bigg(1-\sqrt{1+\frac{128(n-3)e^2}{(n-2)^3\beta^2v^4}}\bigg).
\end{eqnarray}
\indent The critical behavior of NYM and nonlinear electrodynamics black holes are the same in four dimensions. The critical behavior of the BIM black hole in four and higher dimensions are in Refs. \cite{Kubiznak,zou}. In this section, we study the critical behavior of NYM black holes in higher dimensions.
If we consider the spherical case with $k=1$ and impose the conditions \eqref{critical} on the equation \eqref{p1}, we arrive at a cubic equation for the critical points
\begin{eqnarray}\label{p1n}
x^3+px+q=0\,,
\end{eqnarray}
where $ x $, $ p $ and $ q $ are given by
\begin{eqnarray}\label{p1nc}
x=\bigg[v^4_{c}+\frac{128(n-3)e^2}{(n-2)^3\,\beta^2}\bigg] ^{-\frac{1}{2}}\,\,\,\,,\,\,\,
p=-{\frac { 3\left(n-2\right)^3 {\beta}^{2}
}{ 256\,(n-3) \, {e}^{2}}}\,\,\,,\,\,\,
q={\frac {\left( n-2 \right)^5\,{\beta}^{2} }{8192\left( n-3 \right)\,{e}
^{4} }} .
\end{eqnarray}
$ x $ is in terms of $ v_{c} $, therefor to have a positive value for $ v_{c} $, we have the condition
\begin{eqnarray}\label{positivevc}
|x| \leq\frac{(n-2)\,\sqrt{2\,(n-2)(n-3)}\,\beta}{16\,  (n-3)\,e}. 
\end{eqnarray}
\indent To obtain an expression for the critical volume, we have to find the roots of equation (\ref{p1n}). For equation \eqref{p1n}, with $ p<0 $ and real $ q $, one can find one or three real roots. It has three real roots when $ 4{p}^{3}+27{q}^{2}\leq0 $, which leads to
\begin{eqnarray}\label{bet0}
	\beta \geq \beta_{0}=\frac{\sqrt{(n-2)(n-3)}}{4\,e}. 
\end{eqnarray}
\indent We can write the roots in trigonometric form as
\begin{eqnarray}\label{roots}
	x_{k^{'}}=2\,\sqrt {\frac{-p}{3}}\cos \left( \frac{1}{3}\,\arccos \left(\,{\frac {3\,q}{2\,p}
		\sqrt {\frac{3}{\pi}}} \right) -\frac{2\pi k^{'}}{3} \right) 
	\,\,\,\,,\,k^{'}=0,1,2,
\end{eqnarray}
where just $ x_{0} $ and $ x_{1} $ give a physical value for the critical volume $ v_{c} $ in Eq.\eqref{p1nc}. $ x_{0} $ and $ x_{1} $ also satisfy the condition (\ref{positivevc}) which provides
\begin{eqnarray}\label{rangeb}
	\beta_{0}=\frac{\sqrt{(n-2)(n-3)}}{4\,e} \leq \beta \leq \beta_{2},
\end{eqnarray} 
where $ \beta_{2}=\,{\frac {\sqrt {2\,(n-2)(n-3)}}{4\,e}} $. There is also one real root when $ \beta<\beta_{0} $, for which $ v_{c} $ is negative and so the critical point can exist only for $ \beta \geq \beta_{0} $. In the range of $\beta_{0}\leq \beta \leq \beta_{2}$, which we call the 'BI regime', there are two critical points corresponding to $x= x_{0,1}$ in equation \eqref{roots}. Although the critical temperature $ T_{c} $ is positive for both critical points $ v_{c} $, one can find that for $ \beta\geq\beta_{1}=\,\frac{\sqrt {(n-2)(n-3)\left( 6+4\,\sqrt {3} \right)}}{12\,e}$
one of the critical points in the BI regime has negative pressure and thus it is unphysical. We have shown the P-v isotherms in the range of $\beta_{0}\leq\beta\leq\beta_{1} $ and $\beta_{1}\leq\beta\leq\beta_{2} $ for $ n=5 $ in Figs. (\ref{Fig8}) and (\ref{Fig9}), respectively. For values of $\beta$ greater than $\beta_{2}$, there is also one critical point corresponding to $x= x_{1} $ in equation \eqref{roots}. In this range ($ \beta>\beta_{2} $), which we call the 'YM regime', the critical behavior is identical to YM-AdS black hole. In other words, there is just one inflection point for $ T < Tc $ and P-v isotherms are qualitatively identical to that of a Van der Walls fluid. We have depicted the critical behavior in this range ($ \beta>\beta_{2} $) in Fig. (\ref{Fig7}) for $ n=5 $.  For a better understanding, we have brought the above results for $ n=5 $ in Table (\ref{Table1}).\\
\indent We can show that as $\beta \rightarrow \infty $, the critical volume determined from $ x_{1} $ reduces to the critical volume of the YM-AdS black hole. So, we name the branch determined from $x_{1}$ as the YM branch and the branch determined from $x_{0}$ as the BI branch,
\begin{eqnarray}\label{branches}
v_{c}=\bigg[\frac{1}{x^2}-\frac{128\,(n-3)\,e^2}{(n-2)^3\,\beta^2}\bigg] ^{\frac{1}{4}}\,\,,\, x = \left\{
\begin{array}{ll}
$$x_{1},\,\,\,\,\,\,\beta \geq \beta_{0}$$\quad \ {(YM-branch)}\quad &  \\ \\
$$x_{0},\,\,\,\,\,\,\beta \in (\beta_{0},\beta_{2})$$\quad \ {(BI-branch)}.\quad &  
\end{array}
\right.
\end{eqnarray}
\begin{table}[h]
\caption{Critical behavior in BIYM theory in 5 dimensions }
\centering
\begin{tabular}{|l|c|c|c|r|}
\hline	
Nonlinear parameter & \,$ \beta <\sqrt{\frac{3}{8 e}} $\, & \, $ \sqrt{\frac{3}{8 e}}<\beta <\frac{\sqrt{9+6\sqrt{3}}}{6\,e} $ \, & \, $ \frac{\sqrt{9+6\sqrt{3}}}{6\,e} <\beta <\frac{\sqrt{3}}{2\,e} $ \, & \, $ \frac{\sqrt{3}}{2\,e}<\beta $ \, \\
\hline
Number of real roots x & one & two & two & one \, \,\,    \\
Number of critical points & none & two & one & one \,\,  \\
Types of BIYM black hole & Schw & Schw & Schw &\, Schw or RN \,    \\
\hline
\end{tabular}
\label{Table1}
\end{table}
\indent The behaviors of the critical values, $T_{c}$, $v_{c}$ and $P_{c}$ with respect to the nonlinear parameter $\beta$ are depicted in Fig. (\ref{Fig10}) and Fig. (\ref{Fig11}). For large $\beta$, the critical values expand as 
\begin{eqnarray}
v_{c}&=&\frac{4\sqrt{6}\,e}{n-2}-\frac{7\sqrt{6}\, (n-3)}{216e\beta^2}+\mathcal{O}\bigg(\frac{1}{\beta^3}\bigg),\\
T_{c}&=&\frac{\sqrt{6}\,(n-3)}{18\pi e}+\frac{\sqrt{6}\,(n-2)(n-3)^2}{5184e^3\pi\beta^2}+\mathcal{O}\bigg(\frac{1}{\beta^3}\bigg),\\
P_{c}&=&\frac{(n-2)(n-3)}{192\pi e^2}+\frac{7(n-2)^2(n-3)^2}{165888\pi e^4\beta^2}+\mathcal{O}\bigg(\frac{1}{\beta^3}\bigg),\\
\rho_{c}&=&\frac{P_{c}V_{c}}{T_{c}}=\frac{3}{8}-\frac{(n-2)(n-3)}{768 e^2 \beta^2}+\mathcal{O}\bigg(\frac{1}{\beta^3}\bigg).
\end{eqnarray} 
\indent We can observe that in the limit $\beta\rightarrow\infty$, the critical values asymptote to those of the YM-AdS black hole. In this limit, the critical ratio tends to $\rho_{c} \rightarrow 3/8$, independent of the dimension $n$. It is in contrary to the abelian Maxwell theory in which the critical ratio, $\rho_{c}=\frac{P_{c} v_{c}}{T_{c}}$ depends on the dimension $n$ \cite{zou}.
\\
\textbf{Critical behaviors of the ENYM and LNYM solutions}\\
\indent The equations of state for the ENYM and LNYM solutions are defined respectively as
\begin{eqnarray}\label{pEN}
P=\frac{T}{v}-k\frac{(n-3)}{\pi(n-2)v^2}+\frac{\beta^2}{4\pi}\bigg[1-\mathrm{exp}\bigg(-\frac{64(n-3)e^2}{(n-2)^3\beta^2 v^4}\bigg)\bigg],\,\,\,\,\,\,\,\,\,\,\,\,\,\,\,\,\,\,\,\,\,\,\,\,\,\,\,\,\mathrm{EN}
\end{eqnarray} 
and
\begin{eqnarray}\label{pLN}
P=\frac{T}{v}-k\frac{(n-3)}{\pi(n-2)v^2}+\frac{\beta^2}{2\pi}\mathrm{ln}\bigg(1+\frac{32(n-3)e^2}{(n-2)^3\beta^2 v^4}\bigg),\,\,\,\,\,\,\,\,\,\,\,\,\,\,\,\,\,\,\,\,\,\,\,\,\,\,\,\,\,\,\,\,\,\,\,\,\,\,\,\,\,\,\,\,\,\,\,\,\mathrm{LN}.
\end{eqnarray}
\\
\indent The critical behaviors of the ENYM and LNYM black holes are qualitatively the same as the BIYM case. However, the critical points and thus the branches mentioned in relation (\ref{branches}) cannot be determined analytically. Therefore, we solved Eq. \eqref{critical} numerically in order to obtain the critical points. In the following, we will discuss the critical behavior just for the LNYM case.  We have obtained the critical values of the six-dimensional LNYM solutions for different values of $\beta$ in Table (\ref{Table 2}). In 6 dimensions, $\beta_{0}=\frac{\sqrt{3}}{2e}\approx\frac{0.86}{e}$, $\beta_{1}=\frac{\sqrt{6(3+2\sqrt{3})}}{6e}\approx\frac{1.03}{e}$ and $\beta_{2}=\frac{\sqrt{6}}{2e}\approx\frac{1.22}{e} $. The P-v isotherms for $n=6$ and $ e=k=1 $ are displayed in Figs. (\ref{Fig12a}) and (\ref{Fig12b}). In the range of $\beta_{0}\leq \beta \leq \beta_{2}$, there are two critical points as we see in  Fig. (\ref{Fig12a}), but for $\beta > \beta_{2}$, there is only one critical point in Fig. (\ref{Fig12b}) and the P-v isotherm is analogous to the Van der Waals fluid.
\begin{table}[ht]
\centering
\begin{tabular}{|l|c|c|c|c|c|c|c|r|}
\hline	
\,\,\,$ \beta $\, & \,$v_{c1} $\, & \,$v_{c2} $\, &\, $ T_{c1} $ \, &\, $ T_{c2} $ \, &  $ P_{c1}$ \, & $ P_{c2}$ \, & $\rho_{c1} $ & \,$\rho_{c2} $\,\, \\
\hline
\,0.9\, & \, 2.2598 \, & \, 1.2866 \, & \,0.1340\, & \,0.1035 \, & \,0.0214 \, & \,0.0028 \, & \,0.3608 \, & \,0.0350 \, \\    \,\,\,1\, & \, 2.3052 \, & \, 1.1734 \, & \,0.1331\, & \,0.0769 \, & \,0.0210 \, & \,-0.0150 \, & \,0.3645 \, & \,-0.2295 \,\\  \,\,\,5 & \, 2.4446 \, & \, 0.4079 \, & \,0.1300\, & \,-3.2717 \, & \,0.0199 \, & \,-4.8685 \, & \,0.3746 \, & \,0.6070 \,\\
\,\,\,10\, & \,2.4482\,& \,0.2777\,& \,0.1299\, & \,-10.9403 \, & \,0.0199 \, & \,22.4528 \,& 0.3749 \, & 0.5699 \,\\ 
\hline
\end{tabular}
\caption{Critical values of the six-dimensional LNYM black hole for different values of $\beta$. We have set $ e=k=1 $. As $\beta$ increases, the second critical volume $ v_{c2} $ decreases and for $\beta\rightarrow\infty$, the second critical volume disappears and the first one approaches to the critical values of 6D YM-AdS black hole ($ T_{c,YM}=\sqrt{6}/6 \pi e\approx 0.1300 $, $v_{c,YM}=\sqrt{6} e\approx2.4494 $, $P_{c,YM}=1/16 \pi e^2 \approx0.0199$ and $ \rho_{c,YM}=\frac{3}{8}\approx0.375 $). }
\label{Table 2}
\end{table}
\\
\subsection{Gibbs Energy}
\begin{figure}
\centering
\includegraphics[scale=0.35]{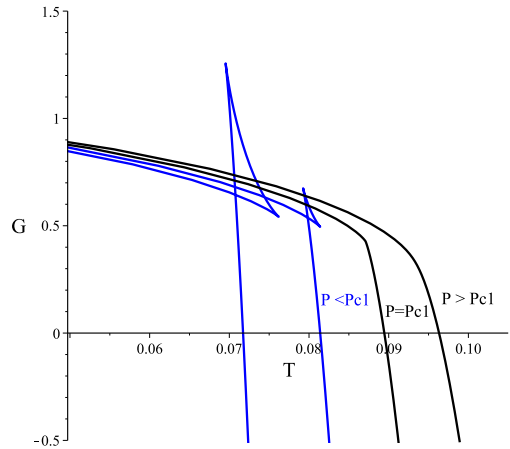}
\caption{\small{\,$\beta > \beta_{2}$. We have set $\beta=2$ and $ e=1 $. For $\beta > \beta_{2}$ the behavior is the same as YM-AdS black hole. There is only one critical point and the corresponding phase transition occurs for $ P<P_{c1} $.} \label{Fig13}}
\end{figure}

\begin{figure}
\centering
\subfigure[\,$\beta\in(\beta_{0},\beta_{1})$. we have set $\beta=0.7$ and $ e=1 $.]{\includegraphics[scale=0.32]{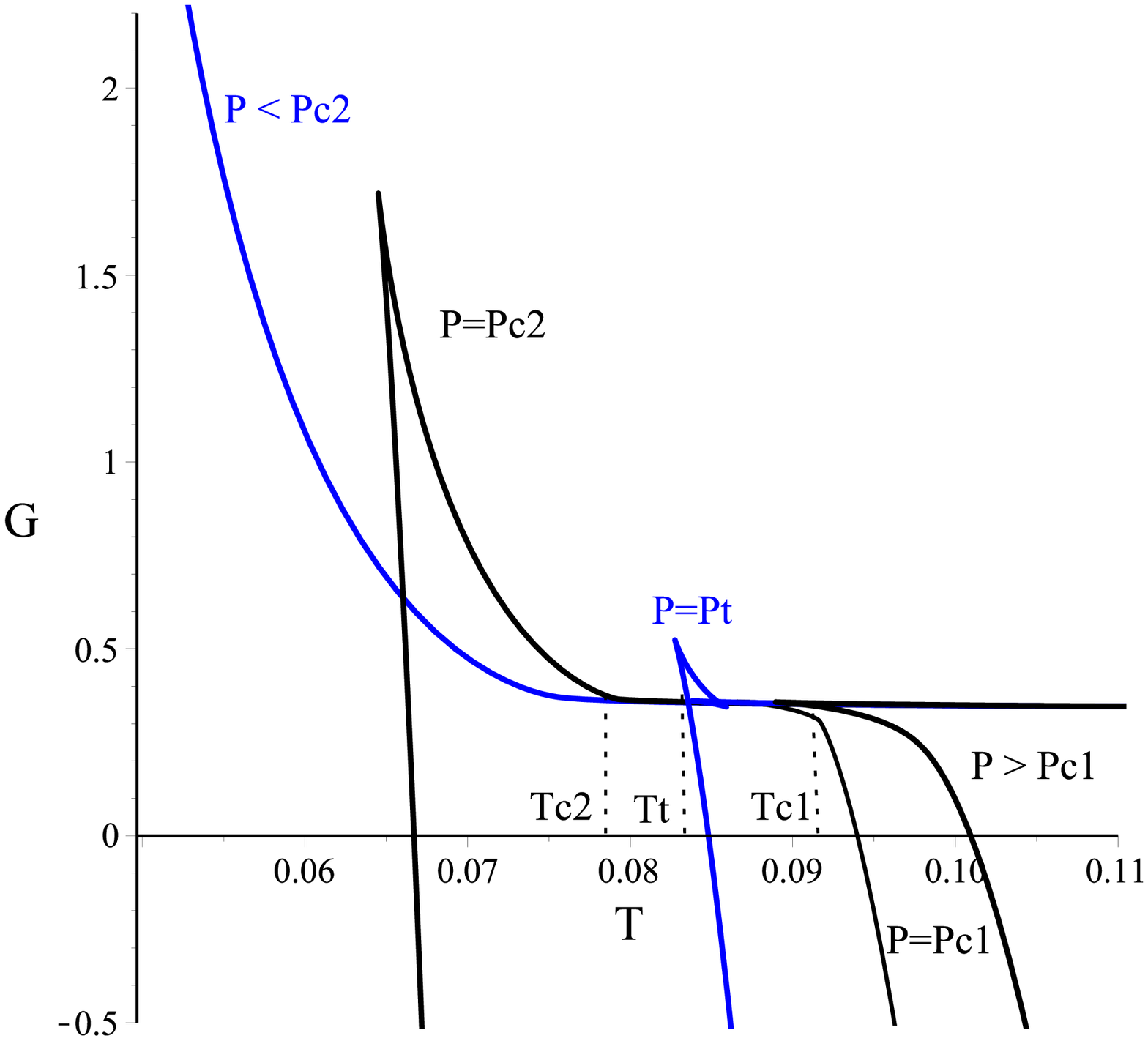}\label{Fig14a}}\hspace*{.5cm}
\subfigure[\,$\beta\in(\beta_{1},\beta_{2}) $. We have set $\beta=0.8$  and $ e=1 $.]{\includegraphics[scale=0.32]{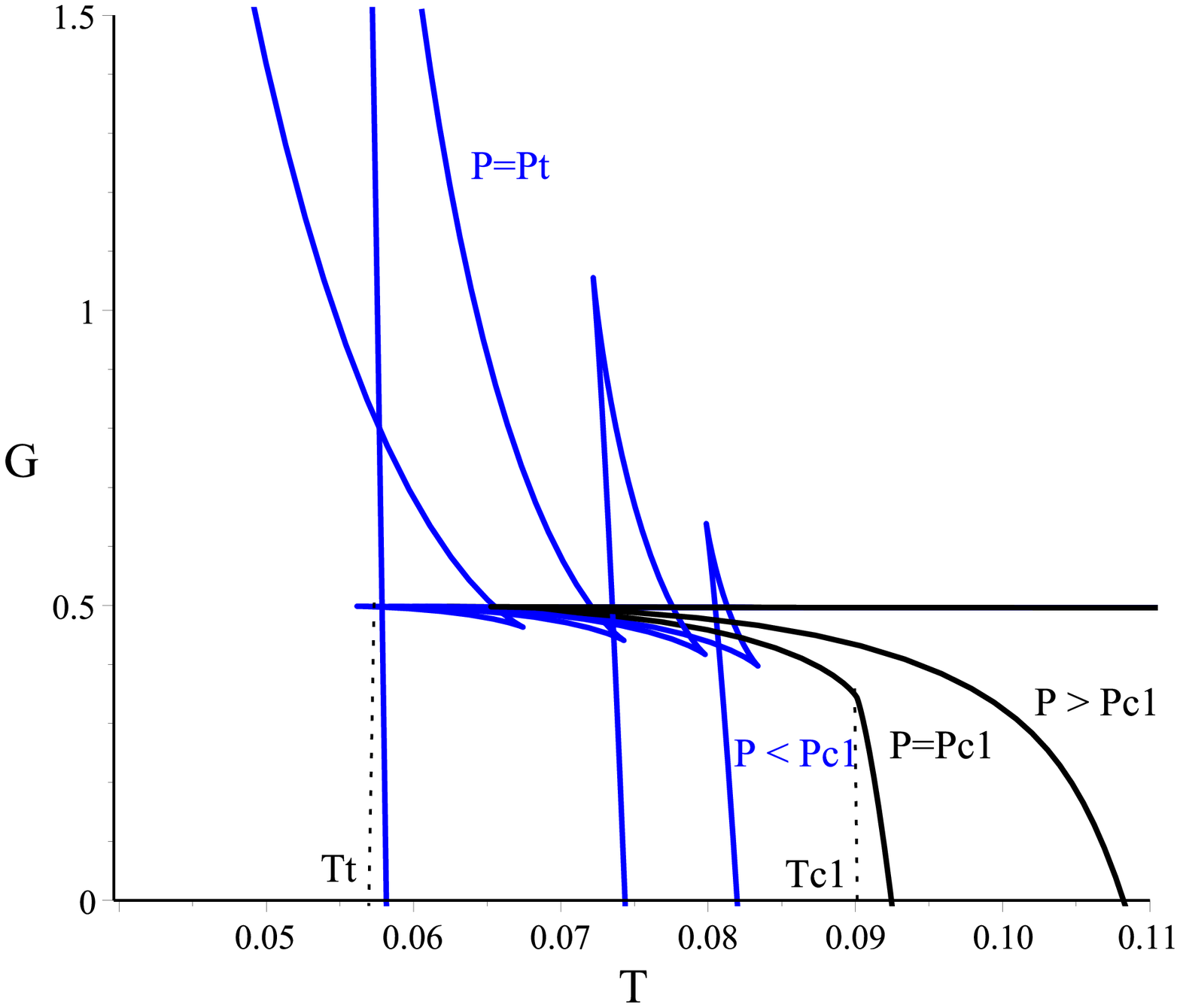}\label{Fig14b}}\caption{ G-T diagram of BIYM theory in $ n=5 $. We have set $ e=k=1 $. In the range of $\beta_{0}\leq\beta\leq\beta_{1}$, there are two critical points with positive pressure. The Gibbs energy is not minimized at $ P=P_{c2} $, and so the first order phase transition only occurs at $ P=P_{c1} $. There is also a reentrant LBH/SBH/LBH transition for $P_{t}\leq P\leq P_{z}$. For $\beta\in(\beta_{1},\beta_{2})$, there is only one physical critical point and so a first order small-large phase transition occurs for $P_{t}\leq P\leq P_{c1} $. We also observe a reentrant phase transition for the range of $ P\in (P_{t},P_{z})$.}\label{Fig14}
\end{figure}

To get more information about the phase transition of the NYM black holes, we analyze the behavior of the Gibbs free energy, $G$. It can be achieved from the relation $G(T, P)=M-TS$. We have plotted $G$ versus $T$ for the five-dimensional BIYM solution in Figs. (\ref{Fig13}) and (\ref{Fig14}). We can observe that the Gibbs free energy depends on the value of the nonlinear parameter $\beta$, as in the case of critical behavior. For $\beta>\beta_{2}$, which we called the YM-branch, the phase transition is the same as the one for YM-AdS black holes or the RN-AdS black holes \cite{Kubiznak}. Namely, there is one critical point that corresponds to a phase transition from a small black hole to a large black hole when $P<P_{c1}$, see Fig. (\ref{Fig13}).\\
\indent We have also shown the behavior of $G$ in the BI branch in Figs. (\ref{Fig14a}) and (\ref{Fig14b}). In Fig. (\ref{Fig14a}) with the range of $\beta\in(\beta_{0},\beta_{1})$, there are two physical (with positive pressure) critical points described by $T_{c1}$ and $T_{c2}$. The phase transition at $T=T_{c2}$ is not physical, since the Gibbs energy is not globally minimized at this point. However, there is a first-order phase transition from a small to a large black hole for $T<T_{c1}$ which ends at $ T=T_{t} $. On the other hand, in a specific range of $T\in (T_{t}, T_{z})$, one can observe a discontinuity in the global minimum of Gibbs energy which shows a reentrant LBH/SBH/LBH (large black hole/small black hole/large black hole) phase transition. Interestingly, this is a special significance for the higher-dimensional NYM solutions while there is no reentrant phase transition for the BI RN-AdS black holes in higher dimensions \cite{Kubiznak}.\\
\indent For the range of $\beta_{1}\leq\beta\leq\beta_{2}$ in Fig. (\ref{Fig14b}), there is only one critical point with positive pressure at $T=T_{c1}$ and a first order SBH/LBH phase transition occurs for $ T_{t}\leq T < T_{c1} $. Similar to the previous case $\beta\in(\beta_{0},\beta_{1})$, in a specific range of $T\in (T_{t}, T_{z})$, the global minimum of G is not continuous, which represents a reentrant LBH/SBH/LBH phase transition.\\
\indent For the LNYM and ENYM cases, the Gibbs energy behavior is qualitatively the same as the BIYM case, so we do not probe them.
\\

\subsection{Critical exponents}
Let us now calculate the critical exponents of the NYM black hole. These values can describe the physical quantities' behavior in the vicinity of the critical point. It should be noted that we use just the physical critical point that we named in the previous sections. There is only one physical critical point for each ranges $\beta>\beta_{2}$, $\beta_{0}\leq\beta\leq\beta_{1}$ and $\beta_{1}\leq\beta\leq\beta_{2}$. To calculate the exponent $\alpha$, we should study the behavior of the heat capacity at constant volume. So, we write the entropy $S$ as a function of $T$ and $V$. If we use Eqs. \eqref{entropy} and \eqref{volume}, we can obtain
\begin{eqnarray}
S(T,V)=\frac{1}{4}\omega_{n-2}^{\frac{1}{n-1}}\bigg(V(n-1)\bigg)^{\frac{n-2}{n-1}},
\end{eqnarray} 
where it shows that the entropy is independent of the temperature $T$ and 
\begin{eqnarray}
C_{V}=T\bigg(\frac{\partial S}{\partial T}\bigg)_{V}=0\,\,\Rightarrow\,\,\,\alpha=0.
\end{eqnarray}
With defining the variables $p$, $\nu$ and $\tau$ as below
\begin{eqnarray}
p=\frac{P}{P_{c}}\,\,\,\,\,,\,\,\,\,\,\nu=\frac{v}{v_{c}}\,\,\,\,\,,\,\,\,\,\,\tau=\frac{T}{T_{c}},
\end{eqnarray}
the Eqs. \eqref{p1}, \eqref{pEN} and \eqref{pLN} can be rewritten as
\begin{eqnarray}\label{expo1}
p&=&\frac{\tau}{\nu \rho_{c}}+h(\nu),
\end{eqnarray} 
where  $\rho_{c}=\frac{P_{c}v_{c}}{T_{c}}$ and
\begin{eqnarray}
h(\nu)&=&-k\frac{n-3}{\pi P_{c}(n-2)\nu^2v_{c}^2}\nonumber\\
+&&\frac{1}{P_{c}}\times\left\{
\begin{array}{ll}
$$-\frac{\beta^2}{4\pi}\bigg(1-\sqrt{1+\frac{128(n-3)e^2}{(n-2)^3\beta^2\nu^4 v_{c}^4}}\bigg)$$,\quad\quad\quad\quad\quad\quad\quad  \ {BI}\quad &  \\ \\
$$\frac{\beta^2}{4\pi}\bigg[1-\mathrm{exp}\bigg(-\frac{64(n-3)e^2}{(n-2)^3\beta^2 \nu^4v_{c}^4}\bigg)\bigg]$$,\quad\quad\quad\quad \quad\quad  \ {EN}\quad &  \\ \\
$$\frac{\beta^2}{2\pi}\mathrm{ln}\bigg(1+\frac{32(n-3)e^2}{(n-2)^3\beta^2 \nu^4v_{c}^4}\bigg)$$.\quad\quad\quad\quad\quad\quad\quad\quad\quad  \ {LN}\quad &
\end{array}
\right.
\end{eqnarray}
\indent Now if we expand Eq. \eqref{expo1} near the critical point, $\tau=t+1$ and $\nu=(1+\omega)^{1/z}$, then we can obtain 
\begin{eqnarray}\label{pa}
p=1+At-Btw-Cw^3+\mathcal{O}(tw^2,w ^4),
\end{eqnarray}
where $B=\frac{1}{z\rho_{c}}$, $C=\frac{1}{z^3}\bigg(\frac{1}{\rho_{c}}-\frac{h^{(3)}|_{\nu=1}}{6}\bigg)$ and $ z=3 $ for all dimensions. The derivative of Eq. \eqref{pa} with respect to $\omega$ for $ t<0 $ gets to 
\begin{eqnarray}
dP=-P_{c}(Bt+3Cw^2)dw.
\end{eqnarray} 
\indent By imposing the Maxwell’s equal area law and considering a constant value for the pressure during the transition, we conclude that
\begin{eqnarray}\label{p}
p=1+At-Bt\omega_{l}-C\omega_{l}^3=1+At-Bt\omega_{s}-C\omega_{s}^{3},\nonumber\\
0=\int_{\omega_{l}}^{\omega_{s}}\omega dP,
\end{eqnarray}
where $\omega_{s}$ and $\omega_{l}$ denote the volume of the small and large black holes. The non-trivial solution of Eq. \eqref{p} is obtained only for $B C t<0$ as below
\begin{eqnarray}
w_{s}=-\omega_{l}=\sqrt{-\frac{Bt}{C}}.
\end{eqnarray}
\indent As it is not possible to find an analytic result for the quantity $ C $, so we gather the numeric results of $A$, $B$, and $C$ for the BI case in Table(\ref{Table 3}). The results show that $\eta=V_{c}(\omega_{l}-\omega_{s})=2V_{c}\omega_{l}\propto\sqrt{-t}$ and so we get to the following result 
\begin{eqnarray}
\beta^{'}=\frac{1}{2}.
\end{eqnarray}
\indent We can differentiate of Eq. \eqref{pa} with respect to $V$ and obtain the isothermal compressibility, $\kappa_{T}=-\frac{1}{V}\frac{\partial V}{\partial P}|_{T}\propto-\frac{V_{c}}{BP_{c}}\frac{1}{t}$. This leads to the critical exponent 
\begin{eqnarray}
\gamma=1.
\end{eqnarray} 
The critical isotherm $t=0$ in Eq. \eqref{pa} reduces to $p-1=-C\omega^3$, where it results to the "shape"  
\begin{eqnarray}
\delta=3.
\end{eqnarray}
\indent The obtained results for $\alpha$, $\beta^{'}$, $\gamma$, and $\delta$ indicate that the critical exponents of the NYM black hole are exactly the same as those of the Van der Waals fluid \cite{Kubiz10}.
\begin{table}[ht]
\caption{The parameters A, B and C for different values of $ \beta $ in BIYM theory for $ n=5 $, $ n=6 $ and $ e=1 $.}
\centering
\begin{tabular}{|c|c|c|c|c|c|c|}
\hline
Parameters &
\multicolumn{3}{c|}{$ n=5 $} & \multicolumn{3}{c|}{$ n=6 $}  \\
\hline
$\beta$ & \,\,\,0.9\, & \,\,\,1\, & \,\,\,2\, & \,1.5\, & \,2\, & \,3\, \\
	
 A  & \, 2.7603\, & \, 2.7372\, & \, 2.6813\, & \,2.7274\, &  \,2.6976\, & \,2.6795\, \\
	
 B  & \,-0.9201\, & \,-0.9124\, & \,-0.8937\, & \,-0.9091\, & \,-0.8992\ & \,-0.8931\, \\

 C & \,-0.0386\, & \,-0.0410\, & \,-0.0475\, & \,-0.0421\, & \,-0.0455\, & \,-0.0477\, \\
\hline
\end{tabular}
\label{Table 3}
\end{table}
\subsection{Dynamical stability of the NYM black hole solutions}
In this section, we would like to investigate the dynamical stability of the nonlinear Yang-Mills black hole. Regge and Wheeler were the first who studied the stability by perturbation modes behaviors\cite{Regge}. They decomposed the perturbations of 4-dimensional static and spherically symmetric background into odd-and even-parity sectors under a two dimensional rotation transformation. Dynamical stability of the nonlinear black hole solutions in Einstein gravity has been studied in Ref. \cite{Moreno}. Now we follow this paper to study the dynamical stability of the NYM black holes. We derive the dynamical stability of the four-dimensional NYM solutions. If one substitutes $ \hat{F}\equiv\frac{1}{4}F^2 $ in the Lagrangian \eqref{22} and also define $\mathcal{L}(\hat{F})\equiv-L(\hat{F})/4$, then the Hamiltonian of the NYM Lagrangian is specified by $ \mathcal{H}\equiv2\mathcal{L}_{\hat{F}}\hat{F}-\mathcal{L} $ where $ \mathcal{L}_{\hat{F}}\equiv\frac{\partial\mathcal{L}}{\partial\hat{F}} $ . It is proper to study the dynamical stability in the so called P frame where $ P\equiv\mathcal{L}_{\hat{F}}^2 \hat{F} $. If $\mathcal{H}_{P}$(where $\mathcal{H}_{P}\equiv\frac{\partial\mathcal{H}}{\partial P}$ ) vanishes nowhere outside the horizon, then the solutions are dynamically stable under odd parity perturbations. We obtain
\begin{eqnarray}\label{fhat}
\mathcal{H}_{P}=\left\{
\begin{array}{ll}
		$$\sqrt{1+\frac{2\hat{F}}{\beta^{2}}}$$,\quad\quad\quad\quad\quad\,\,  \ {BI}\quad &  \\ \\
			$$e^{\frac{\hat{F}}{\beta^{2}}}$$,\quad\quad\quad\quad\quad \quad\quad\,  \ {EN}\quad &  \\ \\
			$$1+\frac{\hat{F}}{2\beta^{2}}$$.\quad\quad\quad\quad\quad\,\,  \ {LN}\quad &
		\end{array}
		\right.
\end{eqnarray}
We evaluate $\mathcal{H}_{P}$ in four dimensions. Using Eq.\eqref{gam} in four dimensions, we get to $ \hat{F}=\frac{e^2}{2r^4} $ that results to the positive value for $\mathcal{H}_{P}$. We can conclude that the NYM solutions are dynamically stable under odd type perturbations since $\mathcal{H}_{P}$ vanishes nowhere outside the horizon. For even-type perturbations, the solutions are dynamically unstable if the condition $ \mathcal{H}_{xx}>0 $ is stablished that $ x=\sqrt{-2Q^2P} $ ($Q$ is the Yang-Mills charge in Eq.\eqref{charge}), and $\mathcal{H}_{xx}=\frac{\partial^2\mathcal{H}}{\partial x ^2}$. For the NYM black holes, $ \mathcal{H}_{xx} $ is obtained as follows
\begin{eqnarray}\label{fhat1}
		\mathcal{H}_{xx}=-\frac{1}{Q^2}\times\left\{
		\begin{array}{ll}
			$$(1+\frac{2\hat{F}}{\beta^{2}})^{\frac{3}{2}}$$,\quad\quad\quad\quad\quad\quad\quad\quad\quad\,\,  \ {BI}\quad &  \\ \\
			$$e^{\frac{\hat{F}}{\beta^{2}}}(1-\frac{2\hat{F}}{\beta^{2}})^{-1}$$,\quad\quad\quad\quad\quad \quad\quad\,  \ {EN}\quad &  \\ \\
			$$(1+\frac{\hat{F}}{2\beta^{2}})^{2}(1-\frac{\hat{F}}{2\beta^{2}})^{-1}.$$\quad\quad\quad\quad\,\,  \ {LN}\quad &
		\end{array}
		\right.
	\end{eqnarray}
We note that ENYM and LNYM black holes are unstable under even-type perturbations for respectively the regions $ \beta^{2}<e^2/r^4 $ and $ \beta^2<e^2/4r^4 $. 
\section{concluding results}\label{result} 
 In this paper, we attained a new $n$-dimensional analytic black hole solution for the non-abelian Yang-Mills gauge theory in the presence of three nonlinear Lagrangians in action, Born-Infeld, exponential and logarithmic Lagrangians. Using the Wu-Yang ansatz, we chose a set of gauge potentials with $SO(n-1)$ and $SO(n-2,1)$ gauge symmetric groups and obtained the spherical and hyperbolic solutions. In four dimensions, the nonlinear Yang-Mills (NYM) solutions are similar to the nonlinear electrodynamics black hole solution in Maxwell theory and so the Yasskin theorem is established. We can also conclude that there is a transformation from the non-abelian gauge fields to a set of the abelian ones in $n=4$. However, for $n\neq4$, we could achieve a new set of nonlinear non-abelian Yang-Mills solutions. As expected, the nonlinear Yang-Mills solutions reduce to the Einstein-Yang-Mills ones when $\beta\rightarrow\infty$. \\  
\indent We checked out the physical structure of the NYM black holes and observed that there is an essential singularity located at $r=0$. Based on the behavior of the metric function in the limit $r\rightarrow 0$, we found two different types of solutions for the horizon (Schw-type and RN-type) in $ n=5 $. In some higher dimensions, the marginal mass is negative and so the Schw-type is the only solution. \\
\indent We also calculated the thermodynamic quantities of the NYM black hole and probed the first law of thermodynamics. We analyzed the thermal stability of the solutions in both the canonical and the grand canonical ensembles. The stable regions happen for the NYM-AdS solutions with $k=-1,1$ and for the flat ones with $k=1$. The physically stable regions decrease as the dimension $n$ increases. In the grand canonical ensemble with the mentioned parameters, stability emerges just for $n=4$, and there are no stable regions for $n=6$. \\  
\indent Furthermore, we investigated the critical behavior of the NYM black holes. We could access a Smarr-type formula for the dimensions $n\neq 4z+1$, where $z$ is an integer parameter. For the spherical solution with $k=1$, we got to the exact critical points for the BIYM black hole, while we used a numeric method to obtain the critical points of the ENYM and LNYM black holes. There is an interesting range of $\beta\in(\beta_{0},\beta_{2})$, in which the critical behavior differs. One can find a discontinuity in the Gibbs energy for $\beta\in(\beta_{0},\beta_{2})$, which indicates a reentrant phase transition, and it is known as the zeroth-order phase transition. This reentrant phase transition only occurs in four dimensions for the nonlinear black hole in Maxwell theory\cite{Kubiznak}. However, in the case of NYM black holes, the reentrant phase transition occurs in four and higher dimensions. For $\beta\rightarrow\infty$, the critical ratio goes to a constant value of 3/8, independent of the dimension $n$. This is one of the differences between the NYM black hole and the nonlinear electrodynamics one, in which the critical ratio depends on the dimension $n$\cite{zou}. We also calculated the critical exponents for the NYM black holes and found that the critical exponents are the same as those in the Van der Waals system.\\
We also probed the dynamical stability of the NYM black holes. For odd-type perturbations, the NYM black holes are dynamically stable. Under even-type perturbations, we face instability for the ENYM and LNYM black holes in the regions $ \beta^{2}<e^2/r^4 $ and $ \beta^2<e^2/4r^4 $, respectively.\\
\indent In this paper, we could reach the nonlinear Yang-Mills solutions in higher dimensions. The compactification of higher dimensional Yang-Mills theories is a very interesting subject. We also expect the compactified Yang-Mills theories to be useful and helpful to better understand phenomenological and theoretical aspects of fundamental physics. We hope to study gravitational aspects of the compactified Yang-Mills theories in the future.
\acknowledgments{This work is supported by Isfahan University of Technology (IUT).}
      
\section{APPENDIX}
\subsection{The Gauge potentials for $SO(3)$, $SO(2,1)$, $SO(4)$ and $SO(3,1)$ gauge groups}\label{app}
The structure constants and the gauge potentials for some gauge groups are defined as follows:\\
For $SO(3)$ gauge group with $k=1$ and $n=4$, the coupling constants and the gauge potentials are defined as
\begin{eqnarray}
	C^{1}_{23}&=& C^{2}_{31}=C^{3}_{12}=-1\,\,,\,\, \gamma_{ab}=\mathrm{diag}(1,1,1)
\end{eqnarray}
and
\begin{eqnarray}\label{aa}
	A_{\mu}^{(i)}&=& A_{\theta}^{(i)} \,d\theta+A_{\phi}^{(i)} \,d\phi\,\,, i=1,2,3,
\end{eqnarray}
where
\begin{eqnarray}
	\left[
	\begin{array}{ccc}
		A_{\mu}^{(1)}\\
		A_{\mu}^{(2)}\\
		A_{\mu}^{(3)} 
	\end{array} \right]&=&e\left[
	\begin{array}{ccc}
		-\,\mathrm{cos}\, \phi & \mathrm{sin}\, \theta \,\mathrm{cos}\,\theta \,\mathrm{sin}\,\phi\\
		-\,\mathrm{sin}\, \phi & -\mathrm{sin}\, \theta \,\mathrm{cos}\,\theta \,\mathrm{cos}\,\phi\\
		0 & \mathrm{sin}^{2}\, \theta
	\end{array} \right]\left[
\begin{array}{ccc}
d\theta \\
\\
d\phi
\end{array} \right].
\end{eqnarray}
For $SO(2,1)$ gauge group with $k=-1$ and $n=4$, the coupling constants are
\begin{eqnarray}
	C^{1}_{23}&=& C^{2}_{31}=-C^{3}_{12}=1\,\,,\,\,\gamma_{ab}=\mathrm{diag}(-1,-1,1),
\end{eqnarray}
where the gauge potentials are followed by Eq. \eqref{aa} with 
\begin{eqnarray}
	\left[
	\begin{array}{ccc}
		A_{\mu}^{(1)}\\
		A_{\mu}^{(2)}\\
		A_{\mu}^{(3)} 
	\end{array} \right]&=&e\left[
	\begin{array}{ccc}
		-\,\mathrm{cos}\, \phi\, & \mathrm{sinh}\, \theta \,\mathrm{cosh}\,\theta \,\mathrm{sin}\,\phi\,\\
		-\,\mathrm{sin}\, \phi\, & -\mathrm{sinh}\, \theta \,\mathrm{cosh}\,\theta \,\mathrm{cos}\,\phi\,\\
		0 & \mathrm{sinh}^{2}\, \theta\,
	\end{array} \right]\left[
\begin{array}{ccc}
d\theta \\
\\
d\phi 
\end{array} \right].
\end{eqnarray}
\indent The coupling constants for $SO(4)$ gauge group with $k=1$ and $n=5$ are described as
\begin{eqnarray}
	C^{1}_{24}&=& C^{1}_{35}=C^{2}_{41}=C^{2}_{36}=C^{3}_{51}=C^{3}_{62}=1,\nonumber\\
	C^{4}_{56}&=& -C^{4}_{21}=C^{5}_{64}=-C^{5}_{31}=C^{6}_{45}=-C^{6}_{32}=1,\nonumber\\
	\gamma_{ab}&=&\mathrm{diag}(1,1,1,1,1,1),
\end{eqnarray}
where the gauge potentials are
\begin{eqnarray}\label{aaa}
	A_{\mu}^{(i)}&=& A_{\theta}^{(i)} \,d\theta+A_{\phi}^{(i)} \,d\phi+A_{\psi}^{(i)} \,d\psi\,\,, i={1,2,3,4,5,6},
\end{eqnarray}
with definitions
\begin{eqnarray}
	\left[
	\begin{array}{ccc}
		A_{\mu}^{(1)}\\
		A_{\mu}^{(2)}\\
		A_{\mu}^{(3)}\\ 
		A_{\mu}^{(4)}\\ 
		A_{\mu}^{(5)}\\ 
		A_{\mu}^{(6)}
	\end{array} \right]&=&e\left[
	\begin{array}{ccc}
		-\,\mathrm{sin}\, \phi\,\mathrm{cos}\, \psi \,& -\mathrm{sin}\, \theta\, \mathrm{cos}\, \theta\, \mathrm{cos}\, \phi \,\mathrm{cos}\,\psi \,& \mathrm{sin}\, \theta\, \mathrm{cos}\, \theta\, \mathrm{sin}\, \phi \,\mathrm{sin}\,\psi\,\\
		-\,\mathrm{sin}\, \phi\,\mathrm{sin}\, \psi \,& -\mathrm{sin}\, \theta\, \mathrm{cos}\, \theta\, \mathrm{cos}\, \phi\, \mathrm{sin}\,\psi \,& -\mathrm{sin}\, \theta\, \mathrm{cos}\, \theta\, \mathrm{sin}\, \phi \,\mathrm{cos}\,\psi\,\\
		-\,\mathrm{cos}\, \phi \,& \mathrm{sin}\, \theta\, \mathrm{cos}\, \theta\, \mathrm{sin}\, \phi \,& 0\\
		0& 0& -\mathrm{sin}^{2}\, \theta\, \mathrm{sin}^{2}\, \phi\, \\
		0&\,\mathrm{sin}^{2}\, \theta\,\mathrm{cos}\, \psi\, & -\mathrm{sin}^{2}\, \theta\, \mathrm{sin}\, \phi\, \mathrm{cos}\, \phi\, \,\mathrm{sin}\,\psi\, \\
		0&\,\mathrm{sin}^{2}\, \theta\,\mathrm{sin}\, \psi\, & \mathrm{sin}^{2}\, \theta\, \mathrm{sin}\, \phi\, \mathrm{cos}\, \phi\, \,\mathrm{cos}\,\psi\, \\
	\end{array} \right]\left[
\begin{array}{ccc}
d\theta \\
\\
d\phi\\
\\
d\psi 
\end{array} \right].
\end{eqnarray}
\indent For $SO(3,1)$ gauge group with $k=-1$ and $n=5$, we have 
\begin{eqnarray}
	C^{1}_{24}&=& C^{1}_{35}=C^{2}_{41}=C^{2}_{36}=C^{3}_{51}=C^{3}_{62}=1\nonumber\\
	C^{4}_{56}&=& C^{4}_{21}=C^{5}_{64}=C^{5}_{31}=C^{6}_{45}=C^{6}_{32}=1\nonumber\\
	\gamma_{ab}&=&\mathrm{diag}(-1,-1,-1,1,1,1),
\end{eqnarray}
and
\begin{eqnarray}
	\left[
	\begin{array}{ccc}
		A_{\mu}^{(1)}\\
		A_{\mu}^{(2)}\\
		A_{\mu}^{(3)}\\ 
		A_{\mu}^{(4)}\\ 
		A_{\mu}^{(5)}\\ 
		A_{\mu}^{(6)}
	\end{array} \right]&=&e\left[
	\begin{array}{ccc}
		-\,\mathrm{sin}\, \phi\,\mathrm{cos}\, \psi\, & -\mathrm{sinh}\, \theta\, \mathrm{cosh}\, \theta\, \mathrm{cos}\, \phi\, \,\mathrm{cos}\,\psi \,& \mathrm{sinh}\, \theta\, \mathrm{cosh}\, \theta\, \mathrm{sin}\, \phi \,\mathrm{sin}\,\psi\,\\
		-\,\mathrm{sin}\, \phi\,\mathrm{sin}\, \psi\, & -\mathrm{sinh}\, \theta\, \mathrm{cosh}\, \theta\, \mathrm{cos}\, \phi\, \mathrm{sin}\,\psi \,& -\mathrm{sinh}\, \theta\, \mathrm{cosh}\, \theta\, \mathrm{sin}\, \phi \,\mathrm{cos}\,\psi\,\\
		-\,\mathrm{cos}\, \phi\, & \mathrm{sinh}\, \theta\, \mathrm{cosh}\, \theta\, \mathrm{sin}\, \phi\, & 0\\
		0& 0& \mathrm{sinh}^{2}\, \theta\, \mathrm{sin}^{2}\, \phi\, \\
		0&-\,\mathrm{sinh}^{2}\, \theta\,\mathrm{cos}\, \psi\, & \mathrm{sinh}^{2}\, \theta\, \mathrm{sin}\, \phi\, \mathrm{cos}\, \phi \,\mathrm{sin}\,\psi\, \\
		0&-\,\mathrm{sinh}^{2}\, \theta\,\mathrm{sin}\, \psi\, &- \mathrm{sinh}^{2}\, \theta\, \mathrm{sin}\, \phi\, \mathrm{cos}\, \phi \,\mathrm{cos}\,\psi\, \\
	\end{array} \right]\left[
\begin{array}{ccc}
d\theta \\
\\
d\phi\\
\\
d\psi 
\end{array} \right].
\end{eqnarray}
where the gauge potentials are followed by Eq. \eqref{aaa}.
\subsection{The metric function $f(r)$ for $n=5$ and $n=9$}\label{f1}
For $n=5$, the solution $f(r)$ in Eq. \eqref{fff} is obtained as follows
\begin{eqnarray}\label{f5}
f(r)&=& k-\frac{m}{r^{2}}-\frac{\Lambda r^2}{6}\nonumber\\
+&&\left\{
\begin{array}{ll}
$$\frac{\beta^2 r^2}{3}\big[1-\sqrt{1+\frac{\eta}{2}}\big]-\frac{e^2}{r^2}\big(\mathrm{ln}[\frac{r^2}{2}(1+\sqrt{1+\frac{\eta}{2}})]-\frac{1}{2}\big)$$,\quad\quad\quad  \ {BI}\quad &  \\ \\
$$-\frac{\beta^2 r^2}{3}\big[1-\mathrm {exp}\big(-\frac{\eta}{4}\big)\big]-\frac{e^2}{2 r^2}\big[E_{i}(1,\frac{\eta}{4})-1+\mathrm{ln}\big(\frac{3e^2}{2\beta^2}\big)+\gamma\big]$$,\quad\quad\quad \quad\quad  \ {EN}\quad &  \\ \\
$$\frac{e^2}{2 r^2}[1-4\,\mathrm{ln}(r)]-\frac{2\beta^2 r^2}{3}(1+\frac{\eta}{8})\mathrm{ln}(1+\frac{\eta}{8})$$,\quad\quad\quad\quad\quad  \ {LN}\quad &
\end{array}
\right.
\end{eqnarray}
where $E_{i}(a,z)=z^{a-1}\Gamma(1-a,z)$ and $\Gamma(a,x)$ and $\gamma$ are respectively the gamma function and  Euler-Mascheroni constant.\\ 
\indent The solution for $n=9$ is
\begin{eqnarray}\label{f9}
f(r)&=& k-\frac{m}{r^{6}}-\frac{\Lambda r^2}{28}\nonumber\\
+&&\left\{
\begin{array}{ll}
$$\frac{\beta^2 r^2}{14}\big[1-\sqrt{1+\frac{21\eta}{6}}\big]-\frac{3e^2}{4r^2}\sqrt{1+\frac{21\eta}{6}}+\frac{21\eta e^2}{8r^2}\big(\mathrm{ln}[\frac{r^2}{2}(1+\sqrt{1+\frac{21\eta}{6}})]+\frac{1}{4}\big)$$,\quad\quad\quad  \ {BI}\quad &  \\ \\
$$-\frac{\beta^2 r^2}{14}\big[1-\big(1-\frac{7\eta}{8}\big)\mathrm {exp}\big(-\frac{7\eta}{8}\big)\big]+\frac{21\eta e^2}{16 r^2}\big[E_{i}(1,\frac{7\eta}{4})-\frac{3}{2}+\mathrm{ln}\big(\frac{21e^2}{2\beta^2}\big)+\gamma\big]$$,\quad\quad\quad \quad\quad  \ {EN}\quad &  \\ \\
$$-\frac{21\eta e^2}{64r^2}(1-8\mathrm{ln}(r))-\frac{3e^2}{4 r^2}-\frac{\beta^2 r^2}{7}\big[1-\big(\frac{49\eta^2}{64}\big)\big]\mathrm{ln}(1+\frac{7\eta}{8})$$.\quad\quad\quad\quad\quad  \ {LN}\quad &
\end{array}
\right.
\end{eqnarray}
\indent It should be noted that we have determined the integration constants in Eq. \eqref{fff} for $ n=5 $ and $ n=9 $ by assuming correspondence of BIYM, ENYM and LNYM theories with Yang-Mills theory for large values of $\beta$.



\begin{thebibliography}{99}
\bibitem{Born}  M. Born and L. Infeld, {\emph {``Foundations of the new field theory"}}, Proc. R. Soc. A \textbf{144}, 425 (1934).

\bibitem{Fradkin} E. S. Fradkin and A. A. Tseytlin, {\emph{"Non-linear electrodynamics from quantized strings"}}, Phys. Lett. B \textbf{163}, 123 (1985).
\bibitem{Tseytlin} A. A. Tseytlin,{\emph{``Vector field effective action in the open superstring theory"}}, Nucl. Phys. B \textbf{276}, 391 (1986).

\bibitem{Abouelsaood} A. Abouelsaood, C. G., Callan, C. R. Nappi and S. A. Yost,{\emph{``Open strings in background gauge fields"}}, Nucl. Phys. B \textbf{280} 599 (1987).

\bibitem{Bergshoeff} E. Bergshoeff, E. Sezgin, C. N. Pope and P. K. Townsend, {\emph{``The Born-Infeld action from conformal invariance of the open superstring"}}, Phys. Lett. B \textbf{188} 70 (1987).

\bibitem{Leigh} R. G. Leigh,{\emph{``Dirac-Born-Infeld action from Dirichlet $\sigma$-model"}}, Mod. Phys. Lett. A \textbf{4}, 2767 (1989).

\bibitem{Heisenberg} W. Heisenberg and H. Euler, {\emph{``Folgerungen aus der Diracschen Theorie des Positrons,"}} Z.
Phys. \textbf{98} 714 (1936);
Translated in En: Consequences of Diracs Theory of the Positron, by W. Korolevski and H.
Kleinert, arXiv:physics/0605038.

\bibitem{Goenner} H. Goenner. \emph{``On the History of Unified Field Theories. Part II."} Living Rev. Relativity \textbf{17} 5 (2014).

\bibitem{Boi} G. Boillat. {\emph{``Nonlinear Electrodynamics: Lagrangians and Equations of Motion"}}, J. Math. Phys. \textbf{11}, 941 (1970). \emph{"Shock Relations in Nonlinear Electrodynamics"}, Phys. Lett. A. \textbf{40} 1 (1972).

\bibitem{Beato} E. Ayón-Beato and A. Garcıa, {\emph{``The Bardeen Model as a Nonlinear Magnetic Monopole"}}, Phys. Lett. B \textbf{493} 149 (2000), arXiv:gr-qc/0009077.

\bibitem{Bandaos} M. Banados and P. G. Ferreira,{\emph{``Eddington's theory of gravity and its progeny"}}, Phys. Rev. Lett \textbf{105} 011101 (2010), arXiv:1006.1769.

\bibitem{Sheykhi0} A. Sheykhi, D. H. Asl, A. Dehyadegari, {\emph{``Conductivity of higher dimensional holographic superconductors with nonlinear electrodynamics"}}, Phys. Lett. B \textbf{781}, 139 (2018).

\bibitem{LN} S. H. Hendi, {\emph{``Asymptotic charged BTZ black hole solutions"}}, J. High Energy Phys. \textbf{03} 065 (2012), arXiv:1405.4941; S. H. Hendi and A. Sheykhi,{\emph{``Charge rotating black string in gravitating nonlinear electromagnetic fields"}}, Phys. Rev. D \textbf{88}, 044044 (2013), arXiv:1405.6998.

\bibitem{EN} H. H. Soleng,{\emph{``Charged black points in General Relativity coupled to the logarithmic U(1) gauge theory"}}, Phys. Rev. D \textbf{52}, 6178 (1995), arXiv:hep-th/9509033.

\bibitem{soo} A. Dehghani, M. R. Setare, and S. Zarepour, {\emph{``Self-Energy Problem, Vacuum Polarization, and Dual Symmetry in Born-Infeld type $U(1)$ Gauge Theories,"}} (2021) arXiv:2112.03757.

\bibitem{Alam} Y. F. Alam, A. Behne,{\emph{``Review of Born-Infeld electrodynamics,"}}, arXiv:2111.08657. 

\bibitem{Soro} D. P. Sorokin, {\emph{``Introductory Notes on Non-linear Electrodynamics and its Applications,"}} arXiv:2112.12118.

\bibitem{Yoko}  N. Yokoi, M. Ishihara, K. Sato and E. Saitoh, {\emph{``Holographic realization of ferromagnets"}},
Phys. Rev. D \textbf{93}, 026002 (2016); https://arxiv.org/abs/1508.01626{}{}{arXiv:1508.01626}.

\bibitem{Hoo} G. Hooft, {\emph{``Topology of the gauge condition and new confinement phases in non-Abelian gauge theories"}}, Nuclear Physics: B 190.3 (1981): 455-478.

\bibitem{Ezawa} Z. F. Ezawa, A. Iwazaki, {\emph{``Abelian dominance and quark confinement in Yang-Mills theories. II. Oblique confinement and $\eta^{'}$ mass"}}, Physical Review D 26.3 (1982): 631.

\bibitem{Kondo} K. I. Kondo, A. Shibata, T. Shinohara and S. Kato, {\emph{``Non-Abelian dual superconductor picture for quark confinement"}}, Phys. Rev. D \textbf{83}, 114016 (2011); {arXiv:1007.2696}.

\bibitem{Nayak} C. Nayak, S. H. Simon, A. Stern, M. Freedman, and S. Das Sarma, {\emph{``Non-Abelian anyons and topological quantum computation"}}, Rev. Mod. Phys. \textbf{80}, 1083 (2008), {arXiv:0707.1889}.

\bibitem{Ivanov} D. A. Ivanov, {\emph{``Non-abelian statistics of half-quantum vortices in p-wave superconductors"}}, Phys. Rev. Lett. \textbf{86}, 268 (2001); https://arxiv.org/abs/cond-mat/0005069v2, {arXiv:cond-mat/0005069}.

\bibitem{Tewari} S. Tewari, S. Das Sarma, C. Nayak, C. Zhang and P. Zoller, {\emph{``Quantum computation using vortices and Majorana zero modes of a $p_{x}+ip_{y}$ superfluid of fermionic cold atoms"}}, Phys. Rev. Lett. \textbf{98} 010506 (2007); https://arxiv.org/abs/quant-ph/0606101v1{}{},\,{arXiv:quant-ph/0606101}.


\bibitem{Bartnik} R. Bartnik and J. McKinnon,{\emph{``Particlelike Solutions of the Einstein-Yang-Mills Equations"}}, Phys. Rev. Lett. \textbf{61}, 141 (1988).

\bibitem{Yasskin} P. B. Yasskin,{\emph{``Solutions for gravity coupled to massless gauge fields"}}, Phys. Rev. D \textbf{12}, 2212 (1975).

\bibitem{Volkov} M. S. Volkov and D. V.Galtsov, JETP Lett. \textbf{50}, 346350, 1989.

\bibitem{bizonp} P. Bizon, Phys. Rev. Lett. 64, 2844, 1990.


\bibitem{Deveci} D. O. Deveciolu, "Lifshitz black holes in Einstein-Yang-Mills theory", Phys. Rev. D \textbf{89} 124020 (2014); arxiv:1401.2133.

\bibitem{Van} J. J. Van der Bij and E. Radu, {\emph{``New hairy black holes with negative cosmological constant"}}, Phys. Lett. B \textbf{536}, 107 (2002), arXiv:gr-qc/0107065.

\bibitem{Okuyama0} N. Okuyama and K. I. Maeda,{\emph{``Domain Wall Dynamics in Brane World and Non-singular Cosmological Models"}}, Phys. Rev. D \textbf{70} 064030 (2004), arXiv:hep-th/0405077.

\bibitem{Okuyama} N. Okuyama and K. I Maeda,{\emph{``Five-dimensional Black Hole and Particle Solution with Non-Abelian Gauge Field"}}, Phys. Rev. D \textbf{67}, 104012 (2003), arXiv:gr-qc/0212022.

\bibitem{Radu} Y. Brihaye, E. Radu, and D. H. Tchrakian,{\emph{``Einstein-Yang-Mills solutions in higher dimensional de Sitter spacetime"}}, Phys. Rev. D \textbf{75}, 024022 (2007), arXiv:gr-qc/0610087.

\bibitem{Brihaye} Y. Brihaye, A. Chakrabarti, B. Hartmann and D. H. Tchrakian,{\emph{``Higher order curvature generalisations of Bartnick-McKinnon and coloured black hole solutions in $d=5$"}}, Phys. Lett. B \textbf{561}, 161 (2003), arXiv:hep-th/0212288.

\bibitem{Wu} T. T. Wu, C. N. Yang and H. Mark, New York, London: Interscience, 349 (1969).

\bibitem{Mazharimousavi} S. H. Mazharimousavi and M. Halilsoy, {\emph{``Einstein-Yang-Mills black hole solution in higher dimensions by the Wu-Yang Ansatz"}}, Phys. Lett. B \textbf{659}, 471 (2008), arXiv:0801.1554.

\bibitem{Dehghani} N. Bostani and M. H. Dehghani,{\emph{``Topological Black Holes of (n+1)-dimensional Einstein-Yang-Mills Gravity"}}, Mod. Phys. Lett. A \textbf{25}, 1507 (2010), arXiv:0908.0661.

\bibitem{Deh00} M. H. Dehghani and A. Bazrafshan, {\emph{``Topological Black Holes of Einstein-Yang-Mills dilaton Gravity"}}, Int. J. Mod. Phys. D \textbf{19}, 293 (2010), arxiv:1005.2387. arXiv:1005.2387
\bibitem{mazhari1}S. H. Mazharimousavi, and M. Halilsoy. "5D black hole solution in Einstein-Yang-Mills-Gauss-Bonnet theory." Physical Review D 76.8 (2007): 087501. arXiv:0801.1562
\bibitem{mazhari2}S. H. Mazharimousavi, and M. Halilsoy, "Higher dimensional Yang–Mills black holes in third order Lovelock gravity." Physics Letters B 665, no. 4 (2008): 125-130. arXiv:0801.1726
\bibitem{mazhari3}S. H. Mazharimousavi, M. Halilsoy, and Zahra Amirabi, "N-dimensional non-abelian dilatonic, stable black holes and their Born–Infeld extension." General Relativity and Gravitation \textbf{42} 261 (2010). arXiv:0802.3990
\bibitem{mazhari4}S. H. Mazharimousavi, and M. Halilsoy. "Black hole solutions in Einstein–Maxwell–Yang–Mills–Gauss–Bonnet theory." Journal of Cosmology and Astroparticle Physics 2008, no. 12 (2008): 005. arXiv:0801.2110

\bibitem{Bostani} M. H. Dehghani, N. Bostani, and R. Pourhasan,{\emph{``Topological Black Holes of Gauss-Bonnet-Yang-Mills Gravity"}}, Int. J. Mod. Phys. D \textbf{19}, 1107 (2010), arXiv:0908.0663. 

\bibitem{Nemati} S. H. Hendi and A. Nemati,{\emph{``Thermodynamics, shadow and quasinormal modes of black holes in five-dimensional Yang-Mills massive gravity"}}, (2019), arXiv:1912.06824.

\bibitem{Stetsko00} M. M. Stetsko, {\emph{``Static spherically symmetric Einstein-Yang-Mills-dilaton black hole and its thermodynamics"}}, Phys. Rev. D \textbf{101}, 124017 (2020), arxiv:2005.13447.

\bibitem{Mir} F. Naeimipour, B. Mirza and F. M. Jahromi, {\emph{``Yang-Mills black holes in quasitopological gravity"}}, The European Physical Journal C \textbf{81}, 13 (2021).


\bibitem{Dyadichev} V. V. Dyadichev and D. V. Gal'tsov,{\emph{``Solitons and black holes in non-Abelian Einstein-Born-Infeld theory"}}, Phys. Lett. B \textbf{486}, 431 (2000), arXiv:hep-th/0005099.

\bibitem{Mazharimousavii} S. H. Mazharimousavi, M. Halilsoy and Z. Amirabi, {\emph{``New non-Abelian black hole solutions in Born-Infeld gravity"}}, Phys. Rev. D \textbf{78} 064050 (2008), arXiv:0806.4614.

\bibitem{Maz} S. H. Mazharimousavi and M. Halilsoy, Lovelock black holes with a power-Yang–Mills source. Physics Letters B, \textbf{681}, 190.(2009). arXiv:0908.0308

\bibitem{Zhang1} M. Zhang, Z. Y. Yang, D. C. Zou, W. Xu, and R. H. Yue, $P-V$ criticality of AdS black hole in the Einstein–Maxwell–power-Yang–Mills gravity. General Relativity and Gravitation, \textbf{47}, 14 (2015). arXiv:1412.1197

\bibitem{Ste1} M. M. Stetsko, Static dilatonic black hole with nonlinear Maxwell and Yang–Mills fields of power-law type. General Relativity and Gravitation, \textbf{53}, 1-21 (2021). arXiv:2012.14915.

\bibitem{Ste2} M. M. Stetsko, {\emph{``Static spherically symmetric black hole in Einstein-power-Yang-Mills-dilaton theory and some aspects of its thermodynamics"}}, arXiv:2012.14902.

\bibitem{Ali} A. Ali and K. Saifullah, . Lovelock black holes surrounded by dark fluid in power-Yang-Mills massive gravity. (2020). arXiv:2006.15610.

\bibitem{Shey3} A. Sheykhi, F. Naeimipour, and S. M. Zebarjad, {\emph{``Phase transition and thermodynamic geometry of topological dilaton black holes in gravitating logarithmic nonlinear electrodynamics"}}, Phys. Rev. D \textbf{91}, 124057 (2015).

\bibitem{Cai0} R. G. Cai, D. W. Pang, and A. Wang, {\emph{``Born-Infeld Black Holes in (A)dS Spaces"}}, Phys. Rev. D \textbf{70}, 124034 (2004), arxiv:hep-th/0410158.

\bibitem{Kubiznak} S. Gunasekaran, D. Kubizňák and R. B. Mann,{\emph{``Extended phase space thermodynamics for charged and rotating black holes and Born-Infeld vacuum polarization"}}, JHEP, \textbf{11}, 110 (2012), arXiv:1208.6251.

\bibitem{zou} Zou, D.C., Zhang, S.J. and Wang, B.. {\emph{``Critical behavior of Born-Infeld AdS black holes in the extended phase space thermodynamics"}}, Phys. Rev. D \textbf{89} 044002 2014, arXiv:1311.7299.

\bibitem{Fernando1}  S. Fernando, {\emph{``Thermodynamics of Born-Infeld-anti-de Sitter black holes in the grand canonical ensemble"}}, Phys. Rev. D \textbf{74} 104 032 (2006),arXiv:hep-th/0608040.

\bibitem{Ams} S. H. Hendi, {\emph{``Asymptotic Reissnes-Nordström black holes"}}, Ann. Phys. (Amsterdam) \textbf{333}, 282 (2013), arXiv:1405.4941.

\bibitem{Malda} J. Maldacena, {\emph{``The Large-N Limit of Superconformal Field Theories and Supergravity"}}, Int. J. Theo. Phys. \textbf{38}, 1113
(1999), arXiv:hep-th/9711200.

\bibitem{Witt} E. Witten, {\emph{``Anti De Sitter Space And Holography"}}, Adv. Theor. Math. Phys. \textbf{2} 253 (1998), arXiv:hep-th/9802150.

\bibitem{Beken} J. D. Beckenstein, {\emph{``Black Holes and Entropy"}},Phys. Rev. D \textbf{7}, 2333 (1973); S. W. Hawking,{\emph{``Black hole explosions?"}}, Nature (London) \textbf{248}, 30 (1974); G. W. Gibbons and S. W. Hawking, Phys. Rev. D \textbf{15}, 2738 (1977).

\bibitem{York}  J. Brown and J. York, {\emph{``Quasilocal energy and conserved charges derived from the gravitational action"}}, Phys. Rev. D \textbf{47}, 1407 (1993).
\bibitem{seyedh} Seyed H. Hendi, S. Panahiyan, and R. Mamasani, "Thermodynamic stability of charged BTZ black holes: Ensemble dependency problem and its solution." General Relativity and Gravitation \textbf{47} 1-24 (2015), arXiv:1507.08496.


\bibitem{Tras} D. Kastor, S. Ray, and J. Traschen, {\emph{``Enthalpy and the mechanics of AdS black holes"}}, Class. Quant. Grav. \textbf{26}, 195011 (2009), arXiv:0904.2765.

\bibitem{Dolan1} B. P. Dolan, {\emph{``Pressure and volume in the first law of black hole thermodynamics}}, Class. Quant. Grav. \textbf{28}, 235017 (2011).


\bibitem{Dolan2} B. Dolan, {\emph{``The cosmological constant and black-hole thermodynamic potentials"}}, Class. Quant. Grav. \textbf{28}, 125020 (2011).

\bibitem{Page1} S. Hawking and D. N. Page,{\emph{``Thermodynamics of black holes in anti-de Sitter space"}}, Commun. Math. Phys. \textbf{87}, 577 (1983).

\bibitem{Kubiz10} D. Kubiznak, R. B. Mann, {\emph{``P-V criticality of charged Ads black holes"}}, JHEP \textbf{07}, 033 (2012), arXiv:1205.0559.

\bibitem{Mirza1} M. B. Jahani Poshteh, B. Mirza, Z. Sherkatghanad, {\emph{``Phase transition, critical behavior, and critical exponents of Myers-Perry black holes"}}, Phys. Rev. D \textbf{88}, 024005 (2013), arXiv:1306.4516.

\bibitem{Kamrani} M. H. Dehghani, S. Kamrani, A. Sheykhi, {\emph{``P-V criticality of charged dilatonic black holes"}}, Phys. Rev. D \textbf{90} 104020,(2014), arXiv:1505.02386.

\bibitem{Mirza2} Z. Sherkatghanad, B. Mirza, Z. Mirzaeyan, S. A. Hosseini Mansoori, {\emph{``Critical behaviors and phase transitions of black holes in higher order gravities and extended phase spaces"}}, Int. J. Mod. Phys. D \textbf{26} 1750017 (2017), arXiv:1412.5028.

\bibitem{Fernando} S. Fernando and D. Krug, {\emph{``Charged Black Hole Solutions in Einstein-Born-Infeld gravity with a Cosmological constant"}}, Gen. Rel. Grav. \textbf{35} 129 (2003), arXiv:hep-th/0306120.


\bibitem{Dayyani} M. H. Dehghani, A. Sheykhi and Z. Dayyani, {\emph{``Critical behavior of Born-Infeld dilaton black holes"}}, Phys. Rev. D \textbf{93}, 024022 (2016), arXiv:1611.08978. 
\bibitem{balartfer} Balart, Leonardo, and Sharmanthie Fernando. "A Smarr formula for charged black holes in nonlinear electrodynamics." Modern Physics Letters A \textbf{32}, 1750219 no. 39 (2017), arXiv:1710.07751 
\bibitem{breton} Bretón, Nora. "Smarr’s formula for black holes with non-linear electrodynamics." General Relativity and Gravitation \textbf{37} 643 (2005). arXiv:gr-qc/0405116
\bibitem{rasheed} Rasheed, D. A. "Non-linear electrodynamics: zeroth and first laws of black hole mechanics." arXiv preprint hep-th/9702087 (1997).

\bibitem{Regge}T. Regge and J. A. Wheeler, "Stability of a Schwarzschild Singularity." Phys. Rev. \textbf{108}, 1063 (1957).
\bibitem{Moreno}C. Moreno and O. Sarbach, "Stability properties of black holes in self-gravitating nonlinear electrodynamics." Phys. Rev. D \textbf{67}, 024028 (2003), arXiv:gr-qc/0208090.

\end{thebibliography}
\end{document}